\documentclass[aps,prd,nofootinbib,twocolumn,superscriptaddress,10pt,floatfix,notitlepage]{revtex4-1}
\pdfoutput=1

\usepackage{soul}
\usepackage[normalem]{ulem}
\usepackage{graphicx}
\usepackage{amsmath,amsfonts,amssymb}
\usepackage[dvipsnames]{xcolor}
\usepackage[breaklinks,colorlinks,urlcolor=blue,citecolor=blue,linkcolor=magenta]{hyperref}
\usepackage{enumitem}
\usepackage{slashed}
\usepackage{bm}
\usepackage{dsfont}
\usepackage{stackrel}
\usepackage{mathrsfs}

\def\be{\begin{equation}}
\def\ee{\end{equation}}

\newcommand{\al}[1]{\begin{align}\begin{aligned} #1 \end{aligned}\end{align}}

\def\bea{\begin{eqnarray}}
\def\eea{\end{eqnarray}}

\newcommand{\braket}[1]{\left\langle #1 \right\rangle}
\newcommand{\sbraket}[1]{\left[ #1 \right]}
\newcommand{\asbraket}[3]{\left\langle #1 \vert #2 \vert #3 \right]}

\newcommand{\bb}[1]{\textbf{#1}}

\newcommand{\dd}{\text{d}}

\newcommand{\del}{\partial}
\newcommand{\La}{\mathscr{L}}
\newcommand{\Amp}{\mathcal{A}}
\newcommand{\cB}{c_{\slashed{B}}}

\usepackage{comment}

\begin{document}
\begin{minipage}{8cm}
\vspace{-1cm}
    \begin{flushright}
DESY-26-058\\
DIAS-STP-26-07
\end{flushright}
\end{minipage}

\title{
Decaying spin-3/2 dark matter from baryon number violation
}

\author{Francesco Costa}
\email{fcosta@stp.dias.ie}
\affiliation{
	School of Theoretical Physics, Dublin Institute for Advanced Studies, 10 Burlington Road, Dublin, D04 C932, Ireland
}

\author{Gabriel M. Salla}
\email{gabriel.massoni.salla@desy.de}
\affiliation{Deutsches Elektronen-Synchrotron DESY, Notkestr. 85, 22607 Hamburg, Germany}

\date{\today}

\begin{abstract}
    We explore an uncharted corner of dark matter phenomenology: non-supersymmetric spin-$3/2$ dark matter with baryon number violating interactions. In an effective field theory description, we identify the leading baryonic portal between the spin-$3/2$ state and Standard Model quarks and show that it can account for the observed dark matter abundance through UV freeze-in and Boltzmann-suppressed freeze-in, while the freeze-out region is completely excluded. The resulting phenomenology is distinctive, with relic production controlled by the competition between baryon-violating single-particle processes and baryon-conserving pair production. We map the viable parameter space against indirect detection, direct detection, and LHC monojet bounds, finding strong complementarity between these probes and especially stringent limits when production and decay are tied to the same operator. We also present a dark QCD-like ultraviolet completion in which the spin-$3/2$ particle arises as a composite baryon, naturally generating the effective interactions and mitigating the main theoretical obstacles of elementary higher-spin states. This framework opens a novel and testable connection between baryonic portals, confining dark sectors, and higher-spin dark matter.
\end{abstract}

\maketitle

%%%%%%%%%%%%%%%%%%%%%%%%%%%%
%%%%% INTRODUCTION %%%%%%%%%
%%%%%%%%%%%%%%%%%%%%%%%%%%%%
\section{Introduction}

Evidence for the existence of dark matter (DM) are well established, ranging from galaxy rotation curves and gravitational lensing to large-scale structure and the cosmic microwave background, all of which point to DM being the most abundant matter component of the universe, cold, non-baryonic and almost collisionless\,\cite{Zwicky:1933gu,Ostriker:1973uit,Planck:2015fie,Planck:2018vyg,Arcadi:2017kky,Clowe:2006eq}. Nevertheless, its microscopic nature remains unknown, i.e. how it is described at the level of fundamental interactions. DM production is often described by two paradigms. In thermal freeze-out, weakly interacting massive particles (WIMPs) are initially in equilibrium with the Standard Model (SM) bath and decouple when their annihilation rate drops below the Hubble expansion rate, fixing the relic abundance\,\cite{Kolb:1990vq}. In contrast, freeze-in assumes negligible initial abundance and extremely feeble interactions, with DM gradually produced from SM processes without ever reaching equilibrium; in this case, the relic density increases with the coupling. This corresponds to feebly interacting massive particles (FIMPs)\,\cite{McDonald:2001vt,Hall:2009bx}. Among the plethora of models that try to address the production of the DM in the early universe, one possible direction that has received comparatively little attention is a DM candidate with spin-3/2 not connected to a supersymmetric framework\,\cite{Yu:2011by,Ding:2012sm,Ding:2013nvx,Chang:2017dvm,Christensen:2013aua,Savvidy:2012qa,Khojali:2016pvu,Khojali:2017tuv,Garcia:2020hyo,Criado:2020jkp}. On the one hand, within supersymmetric theories, in particular supergravity, the gravitino\,\cite{Freedman:1976xh,Deser:1976eh}, the spin-3/2 superpartner of the graviton, emerges as a natural DM candidate, see for instance Refs.\,\cite{Bolz:1998ek,Pagels:1981ke,Moroi:1993mb,Covi:2009bk,Buchmuller:2007ui}. On the other hand, given that so far no evidence for supersymmetry were found at collider experiments, non-supersymmetric alternatives to spin-3/2 particles become increasingly compelling. Such non-supersymmetric spin-3/2 states can emerge as a composite resonance of a strongly interacting dark sector. For instance, as we will discuss in more detail, in one-flavour confining $SU(3)$ theories the lightest and stable baryon carries spin 3/2, therefore making our DM candidate a possible outcome of confinement. A large body of work has explored strongly interacting dark sectors and their rich phenomenology\,\cite{Garani:2021zrr,CMS:2021dzg,Cheng:2021kjg,Strassler:2006im,Cohen:2017pzm,Buckley:2012ky,Kribs:2016cew,Mitridate:2017oky,Lonsdale:2017mzg,Appelquist:2015yfa,Hardy:2014mqa}.

As opposed to lower spin particles, general (non-supersymmetric) spin-3/2 particles are subjected to a number of theoretical constraints. First, it was proved long ago that charged spin-3/2 particles suffer from super-luminal propagation\,\cite{Velo:1969bt,Velo:1969txo}. As a consequence, while it may not be possible to write renormalizable interactions of the spin-3/2 to the SM particles, one can write higher-dimensional interactions in an effective field theory (EFT) framework. Second, it was shown that the presence of a spin-3/2 state in the early universe may lead to a catastrophic production of particles due to the expanding space-time\,\cite{Kolb:2025wyj,Kolb:2021xfn,Kaneta:2023uwi,Qiu:2022lyn,Antoniadis:2021jtg,Castellano:2021yye,Cribiori:2021gbf,Terada:2021rtp,Dudas:2021njv}. Lastly, the consistency of effective interactions of spin-3/2 particles was recently put under scrutiny in Refs.\,\cite{Bellazzini:2025shd,Gherghetta:2025tlx} (see also Ref.\,\cite{Grisaru:1977kk}), where on-shell and positivity techniques were used to argue that elementary spin-3/2 states cannot have well defined effective interactions unless the spin-3/2 particle is the gravitino. These show that models of fundamental non-supersymmetric spin-3/2 particles are theoretically constrained. 
However, issues related to the effective interactions and gravitational particle production are naturally avoided, for instance by considering a composite spin-$3/2$ case, in which this state emerges as the lightest state of a dark QCD-like sector, as we will discuss in detail later on.
Further aspects of higher-spin particles can be found in Refs.\,\cite{Schroer:2017nly,Afkhami-Jeddi:2018apj,Criado:2020jkp,Trott:2020ebl}.

Several works have explored effective interactions of non-supersymmetric spin-3/2 DM. Requiring DM stability, most studies have focused on interactions involving two spin-3/2 fields. Among these, the lowest-dimensional operator is the dimension-five interaction $\bar \Psi_\mu \Psi^\mu |H|^2$, with $\Psi_\mu$ the spin-3/2 and $H$ the SM Higgs fields\,\cite{Chang:2017dvm}. Dimension six interactions involving two SM fermion fields were also analysed\,\cite{Yu:2011by,Ding:2012sm,Ding:2013nvx,Khojali:2016pvu,Khojali:2017tuv}. In Ref.\,\cite{Garcia:2020hyo} instead, lepton number violating (LNV) dimension five interactions involving a single spin-3/2 field were discussed, where particular attention to decays of the DM was given. Furthermore, in Ref.\,\cite{Dong:2024dce} the baryon number conserving Hilbert series of operators up to dimension eight was presented. A point missed in the literature so far is the \textit{baryon number violating} (BNV) interactions of a spin-3/2 particle.

In this paper we will investigate non-supersymmetric spin-3/2 theories that possess BNV interactions. Similar to the case of spin-1/2 particles, we expect the phenomenology and UV structure of LNV\,\cite{Chikashige:1980ui,Schechter:1981cv,Lazarides:2018aev,Chao:2023ojl, Deppisch:2015qwa,Pilaftsis:1991ug,Furry:1939qr,Avignone:2007fu,Bilenky:1987ty,Doi:1985dx} and BNV\,\cite{Weinberg:1979sa,Mohapatra:1980qe,Curtin:2018mvb,Cui:2012jh,Cui:2013bta,Cui:2014twa,Cui:2016rqt,Ipek:2016bpf,Aitken:2017wie,McKeen:2015cuz,Davoudiasl:2010am,Davoudiasl:2015jja,Arnold:2012sd,Assad:2017iib,Cheung:2013hza,Allahverdi:2010rh,Allahverdi:2013mza,Allahverdi:2017edd,Bittar:2024fau,Bittar:2024nrn} theories to be vastly different. More precisely, we will be interested in studying how the DM abundance can be reproduced from such a theory, while taking into account the experimental constraints as well as the theoretical considerations mentioned above.

The paper is organized as follows. In Sec.\,\ref{sec:EFT} we classify the leading effective operators involving 
spin-$3/2$ fields and coupling to the quarks, with particular attention to BNV interactions. We compute the relevant observables to our analysis, emphasising also how the spin-3/2 case differs from the spin-1/2 one. In Sec.\,\ref{sec:DM} we discuss the generation of DM abundance from these couplings in different regimes, presenting both numerical and analytical approximations for the relic density. In Sec.\,\ref{sec:pheno} we analyse phenomenological signatures and detection prospects, which includes DM direct and indirect detection and monojet signals at colliders. In Sec.\,\ref{sec:uv} we discuss a possible UV completion of our model, that is based on confining dark sectors where the spin-$3/2$ state arises as a composite baryon. We also discuss throughout the text how this model may evade the theoretical constraints on spin-3/2 particles, while also briefly mentioning the possibility of achieving baryogenesis within this framework. Finally, we present our conclusions in Sec.\,\ref{sec:conclusions}. We have three appendices: in App.\,\ref{app:spin32} we elaborate more on the theory of spin-3/2 particles, in App.\,\ref{app:misc} we justify the absence of some BNV operators and in App.\,\ref{app:boltzmannEq} more details on the Boltzmann equations are given.
\\

%%%%%%%%%%%%%%%%%%%%%%%%
%%%%% Spin 3/2 EFT %%%%%
%%%%%%%%%%%%%%%%%%%%%%%%

\section{Spin-3/2 EFT}\label{sec:EFT}

A massive spin-3/2 state can be described by a Rarita--Schwinger field, namely a field $\Psi_\mu$ that carries both spinor and vector Lorentz indices\,\cite{Rarita:1941mf}. 
To ensure that only the physical degrees of freedom of the massive spin-$3/2$ particle propagate, one must impose the additional constraints $\gamma^\mu \Psi_\mu = 0$ and $\partial^\mu \Psi_\mu = 0 \,$
which then allows us to write the equations of motion as four identical Dirac equations:
\be
(i\slashed{\del}-m_\Psi)\Psi_\mu=0,
\ee
where $m_\Psi$ is the (Dirac) mass of the particle. We give more details on the theory of the spin-3/2 field in App.\,\ref{app:spin32}.

The interactions between the spin-3/2 and the SM particles take place necessarily at the non-renormalisable level. Assuming the spin-3/2 field to be a complete singlet under the SM gauge group, the first interactions one can write are of dimension five:
\al{\label{eq:La_dim5}
\La^{(5)} = \frac{1}{\Lambda}\bar\Psi_\mu\left(c_{HH}^V + c_{HH}^A\gamma_5\right)\Psi^\mu |H|^2 \\ + \frac{c_{L}}{\Lambda} \bar L D^\mu \tilde H\Psi_\mu ,
}
where $\Lambda$ is the EFT cut-off, $c_{HH}^{V,A},c_{L}$ are dimensionless coefficients, $L$ and $H$ the SM lepton and Higgs $SU(2)_L$ doublets, respectively, and $\tilde H= i\sigma_2H^*$. After electroweak symmetry breaking, the first operators in the equation above can modify the Higgs decay width\,\cite{Chang:2017dvm}, if kinematically allowed, and in addition also induce a cross-section $\sigma(\bar \Psi \Psi\to \text{SM+SM)}$ that can be used to produce the spin-3/2 in the early universe and reproduce the DM abundance via freeze-out\,\cite{Chang:2017dvm}. The production via freeze-in at high and low reheating temperatures with same Higgs portal operator is instead studied in Ref.~\cite{Costa2026}. The second operator instead has been studied in Ref.\,\cite{Garcia:2020hyo}, in particular it can make the spin-3/2 decay to SM particles and also generate the DM abundance observed today via freeze-in.

While the first interaction in Eq.\,\eqref{eq:La_dim5} is in principle always allowed by the SM gauge and global symmetries, the second one explicitly violates lepton number. In order to have this operator present at low energies, one would therefore require a specific UV dynamics that induces such LNV connected to the spin-3/2 state. With similar reasoning, one could consider that the spin-3/2 field violates baryon number instead of lepton number, which would then be mapped to a whole different set of UV completions. To the extent of our knowledge such interactions were not previously considered in the literature; in particular, previous attempts to write down effective interactions of spin-3/2 particles were restricted to LNV interactions only and hence missed these operators\,\cite{Dong:2024dce}. Given that the spin-3/2 field is taken as a SM singlet, the first BNV operator we can write emerges at dimension six and it is
\be\label{eq:La_dim6_BNV}
\La_\text{BNV}^{(6)} = \frac{\cB^{ijk}}{\Lambda^2}(\bar \Psi_\mu Q_i)(\bar Q^c_j\gamma^\mu d_{Rk})+h.c.,
\ee
where $i,j,k$ are flavour indices, $Q$ is the SM quark doublet, $d_R$ the down-type right-handed quark and $Q^c$ denotes the charge conjugate field of $Q$. Due to the fact that the $SU(2)_L$ and colour $SU(3)_c$ indices of the SM fields are contracted by the respective Levi-Civita tensors, we have that $\cB^{ijk}$ is antisymmetric in the $Q\bar Q^c$ flavour indices $\cB^{ijk}=-\cB^{jik}$, implying that at least two generations are involved. Another possible BNV operator at dimension six would be one involving up-type right-handed quarks, namely $u_Rd_Rd_R$, however, we show in App.\,\ref{app:misc} that such operator produces an identically zero on-shell amplitude, rendering the operator in Eq.\,\eqref{eq:La_dim6_BNV} the single relevant BNV operator at dimension six. Such is not the case of spin-1/2 particles, as both operators with $Q\bar Q^cd_R$ and $u_R\bar d_R^c d_R$ are allowed.

Still at the level of dimension six operators, many operators involving two spin-3/2 fields and two SM fermions can be constructed. Those were listed in Ref.\,\cite{Dong:2024dce} and studied in Refs.\,\cite{Yu:2011by,Ding:2012sm,Ding:2013nvx,Savvidy:2012qa,Khojali:2016pvu,Khojali:2017tuv,Criado:2020jkp} in the context of DM and, because of the extra cut-off suppression, typically lead to looser constraints on $\Lambda$ with respect to the dimension five operators. Examples of such operators are
\al{\label{eq:La_dim6_qq}
\La^{(6)} \supset \; &\frac{c_{QQ}^{ij}}{\Lambda^2}(\bar \Psi_\mu Q_i)(\bar Q_j \Psi^\mu) \\ & + \frac{c_{dd}^{kl}}{\Lambda^2}(\bar \Psi_\mu d_{Rk})(\bar d_{Rl} \Psi^\mu)+h.c.,
}
which are baryon number conserving. Here, $i,j,k,l$ are flavour indices. As we will show later on in Sec.\,\ref{sec:uv}, UV completions that generate the BNV operator in Eq.\,\eqref{eq:La_dim6_BNV} are more naturally connected to the SM quarks rather than leptons, so that we can expect operators such as the ones in Eq.\,\eqref{eq:La_dim6_qq} to be particularly relevant. In our specific UV construction the two operators above can be generated at tree-level after integrating out the heavy degrees of freedom.

In what follows, we will focus on studying the role that the BNV operator in Eq.\,\eqref{eq:La_dim6_BNV} can play in producing the entirety of the DM, as well as on the experimental and theoretical constraints we can impose on it. We will also discuss afterwards the operators in Eq.\,\eqref{eq:La_dim6_qq} as a way to understand their interplay and to make contact with the UV completion to be presented in Sec.\,\ref{sec:uv}.

A remark is in order at this point. Given that the operator in Eq.\,\eqref{eq:La_dim6_BNV}  violates baryon number, one might raise the possibility of inducing signatures such as proton decay and baryon-antibaryon oscillations. Regarding proton decay, we will always consider the spin-3/2 particle to have a mass above $2$ GeV, thus kinematically forbidding the decay. Instead, baryon-antibaryon oscillation cannot be generated as the spin-3/2 state is assumed to be Dirac. This can be understood from the fact that we can assign a baryon number charge to $\Psi_\mu$ such that the Lagrangian remains invariant under baryon number symmetry. With the addition of a Majorana mass that explicitly breaks baryon number symmetry and given the antisymmetric flavour structure of $\cB^{ijk}$, one could generate $\Lambda^0\bar\Lambda^0$ oscillation at tree-level and $n\bar n$ at higher-loop oder\,\cite{Bittar:2024nrn,Beneito:2025ond}. Furthermore, one could entertain the idea of generating the observed baryon-antibaryon asymmetry of the universe with the BNV operator\,\cite{Sakharov:1967dj,Cohen:1993nk,Davidson:2008bu}. The Lagrangian $\La^{(6)}_\text{BNV}$ by itself is not enough to achieve successful baryogenesis, as the spin-3/2 particles, assumed Dirac at this point, are produced in a equal quantities as their antiparticles, hence no net baryon asymmetry is created. The first requirement would therefore be to include a Majorana mass to $\Psi_\mu$ that will explicitly break baryon number. Then, we see two main directions in which one could generate the correct asymmetry. First, inspired by models of baryogenesis with spin-1/2 particles, one can enrich the particle spectrum with additional spin-3/2 states and scalars, and demand the spin-3/2 to decay out-of-equilibrium in the early universe. For analogous interactions with spin-1/2 particles, it was shown that the correct baryon asymmetry can be generated in this way\,\cite{Davoudiasl:2010am,Davoudiasl:2015jja,Arnold:2012sd,Assad:2017iib,Cheung:2013hza,Allahverdi:2010rh,Allahverdi:2013mza,Allahverdi:2017edd,Bittar:2024nrn,Bittar:2024fau}. A second approach would be, in a similar spirit to mesogenesis scenarios\,\cite{Aitken:2017wie, Elor:2018twp}, to profit from the fact that the spin-3/2 particle can directly mix with SM baryons, which in turn can decay with the necessary $CP$-asymmetry. The precise model building of these scenarios and the study of the corresponding flavour constraints lies beyond the scope of this paper, hence is left for future work.

%%%%%%%%%%%%%%%%%%%%%%%%%%%%
%%%%% DM abundance %%%%%%%%%
%%%%%%%%%%%%%%%%%%%%%%%%%%%%

\section{DM abundance}\label{sec:DM}

In this Section we will discuss the generation of the spin-$3/2$ DM abundance. As outlined previously, we will focus on production via non-renormalisable interactions with SM quarks from Eqs.\,\eqref{eq:La_dim6_BNV} and \eqref{eq:La_dim6_qq}. The results provided in this Section are trustworthy as long as the EFT is valid, therefore, we will always neglect the regions where $\Lambda < m_\Psi$. Furthermore, for EFT to be a good description of freeze-in production in this model, we will also need $\Lambda > T_\text{RH}$, where $T_\text{RH}$ is the reheating temperature. That is necessary because, for non-renormalisable operators, freeze-in is a UV dominated process and most of the production occurs at $T \sim T_\text{RH}$\,\cite{Elahi:2014fsa}. This discussion is of particular relevance depending on the nature of the spin-3/2 particle, that is, if it is elementary or composite: If the spin-$3/2$ is a composite particle  its production is only well defined for temperatures below the confinement scale. In this Section we will remain agnostic about the UV origin of the spin-3/2, and we will elaborate further on this matter and on UV completions in Sec.\,\ref{sec:uv}. 

\subsection{Boltzmann equation and rates}

The evolution of the spin-$3/2$ number density $n_\Psi$ can be described by Boltzmann equations. Here, we will use the following Boltzmann equation for the number density $n_\Psi$:
\al{\label{eq:Boltzmann_eq_general}
\frac{H S x}{\Delta} \frac{\mathrm{d} Y_\Psi}{\mathrm{d}x} 
& = \dot n_\Psi + 3Hn_\Psi \\ 
& = \; -\gamma_{\Psi  \leftrightarrow qqq}  \frac{Y_{\Psi}}{Y^{\rm eq}_{\Psi}}+\gamma_{\Psi \bar q \leftrightarrow qq} \left(1- \frac{Y_{\Psi}}{Y^{\rm eq}_{\Psi} }   \right) \\
& \hspace{0.5cm}+\; \gamma_{\Psi \bar\Psi \leftrightarrow q\bar q}  \left(1-\frac{Y_{\Psi}^2}{{Y^{\rm eq}_{\Psi}}^2}   \right),
}
where $H,~S$ are the Hubble and entropy density, respectively, and $Y_i=n_i/S$ are the corresponding yields. Also, $x=m_\Psi/T$, $Y^\text{eq}$ are the equilibrium yields and $\Delta=1+\frac{T}{3g_{*s}(T)}\frac{\del g_{*s}(T)}{\del T}$ takes into account the variation of the entropic degrees of freedom $g_{*s}(T)$. In the equation above we have included the 2-to-2 processes $\bar q \Psi\leftrightarrow qq$ and $q\bar q \leftrightarrow \Psi\bar \Psi$, which arise from the Lagrangians in Eqs.\,\eqref{eq:La_dim6_BNV} and \eqref{eq:La_dim6_qq}, respectively, as well as the decay channel $\Psi\leftrightarrow qqq$ that comes solely from the BNV operator. We have assumed that all quarks, denoted collectively by $q$, are in thermal equilibrium. The reaction rates above are
\al{
\gamma_{1+2 \leftrightarrow a+b} 
& = n_1^{\rm eq} n_2^{\rm eq} \langle \sigma_{1+2 \leftrightarrow a+b} v \rangle,\\
\gamma_{1\leftrightarrow a+b,..,j} 
& = n_1^{\rm eq} \langle \Gamma_1\rangle,
}
where $\langle \sigma_{1+2 \leftrightarrow a+b }  \rangle$ and $\langle\Gamma_1\rangle$ are the thermally averaged cross section and decay rate. More details on the Boltzmann equations and the relevant thermal averages expressions are given in the App.\,\ref{app:boltzmannEq}.

The cross-sections and lifetime can be computed starting from Eqs.\,\eqref{eq:La_dim6_BNV} and \eqref{eq:La_dim6_qq}. Assuming massless quarks, the 2-to-2 cross-sections in the center of mass frame for the two operators we considered are
\be\label{eq:sigma_qPsi_qq}
\sigma_{\bar q \Psi\to qq}(s) = N_f(s)|c_{\slashed{B}}|^2\frac{s^2+10m_\Psi^2s+m_\Psi^4}{72\pi m_\Psi^2 \Lambda^4},
\ee
and
\al{\label{eq:sigma_PsiPsi_qq}
\sigma_{\Psi\bar\Psi\to d_R\bar d_R} (s)=&\frac{|c_{dd}|^2N_f'(s)}{6912 \pi m_\Psi^4\Lambda^4} \\ \times &\frac{(s^3-5s^2m_\Psi^2+12m_\Psi^4s-2m_\Psi^6)}{\sqrt{1-4m_\Psi^2/s}},
}
where $s$ is the center of mass energy squared. We consider flavour universal couplings $\cB,c_{dd}$ with all three generations, whereas $N_f$, $N_f'$ take into account the allowed number of flavour combinations for each of the operators. They take values $N_f=N_f'=9$ when all quarks contribute to the scattering and their lowest value is $1$ when only $uds$ quarks are involved. The cross-section in Eq.\,\eqref{eq:sigma_PsiPsi_qq} was computed from the operator involving $d_R\bar d_R$ in the Lagrangian\,\eqref{eq:La_dim6_qq}, while for the cross-section of the operator with $Q\bar Q$ we obtain an identical expression except for an order one numerical factor and a different $N_f'$. Furthermore, we have checked that for other independent operators involving two quarks and two spin-3/2 the dependence on the energy and mass is unchanged and has the same order of magnitude as $\sigma_{\Psi\bar\Psi\to d_R\bar d_R}$. Hence, for our purposes, we take 
\be
\sigma_{\Psi\bar\Psi\to q\bar q}\sim \sigma_{\Psi\bar\Psi\to d_R\bar d_R},
\ee
with a coefficient $c_{qq}$. Notice that both cross-sections in Eqs.\,\eqref{eq:sigma_qPsi_qq} and \eqref{eq:sigma_PsiPsi_qq} are inversely proportional to the mass $m_\Psi$, which is a consequence of the propagation of the longitudinal modes of the spin-3/2 (see Eq.\,\eqref{eq:RS_propagator}).

The decay rate from the BNV operator is given as
\al{\label{eq:width_quarks}
&\Gamma_{\Psi}(\Psi\to QQd_R) = \frac{N_f(m_\Psi)m_\Psi^5|c_{\slashed{B}}|^2}{320\pi^3\Lambda^4},\\
& \qquad \qquad \quad \mathrm{for } \; m_\Psi\gtrsim 2~\text{GeV},
}
whose scaling is what we would expect from dimensional analysis. Considering the channel $\Psi\to QQd_R$ above to be the leading contribution to the total width and given that $\tau_\text{Uni}\simeq 4.3\times 10^{17}$ s, the DM lifetime can be expressed as
\be\label{eq:life-time}
\frac{\tau_\Psi}{\tau_\text{Uni}} \simeq 1.2\left(\frac{4}{N_f}\right)\left(\frac{2~\text{GeV}}{m_\Psi}\right)^5\left(\frac{\Lambda/\sqrt{|c_{\slashed{B}}|}}{10^{10}\text{GeV}}\right)^4.
\ee
Notice that in Eq.\,\eqref{eq:width_quarks} we highlighted that we should take the value of the width to quarks only if the spin-3/2 mass is above the GeV scale. If not, the relevant degrees of freedom are the hadrons and we need to compute the width to hadronic final states, rendering the computation more challenging. Moreover, given that $\Psi$ has baryon number one, it must decay to a final state with net baryon number one as well, e.g. $p^+\pi^-$, $n\pi^0,\cdots$. This in turn implies that for light enough $\Psi$, one can have decay channels for the proton such as $p^+\to \Psi \pi^+,\cdots$. Thus, in order to avoid proton decay and at the same time to use Eq.\,\eqref{eq:width_quarks} consistently, we always take $m_\Psi\gtrsim 2$ GeV.

In what follows, we will first obtain approximate analytical expressions for the DM abundance in the freeze-in and freeze-out regimes, and afterwards compare them with the numerical results.

\subsection{Freeze-in results}

\begin{figure*}
    \centering
    \includegraphics[width=0.48\linewidth]{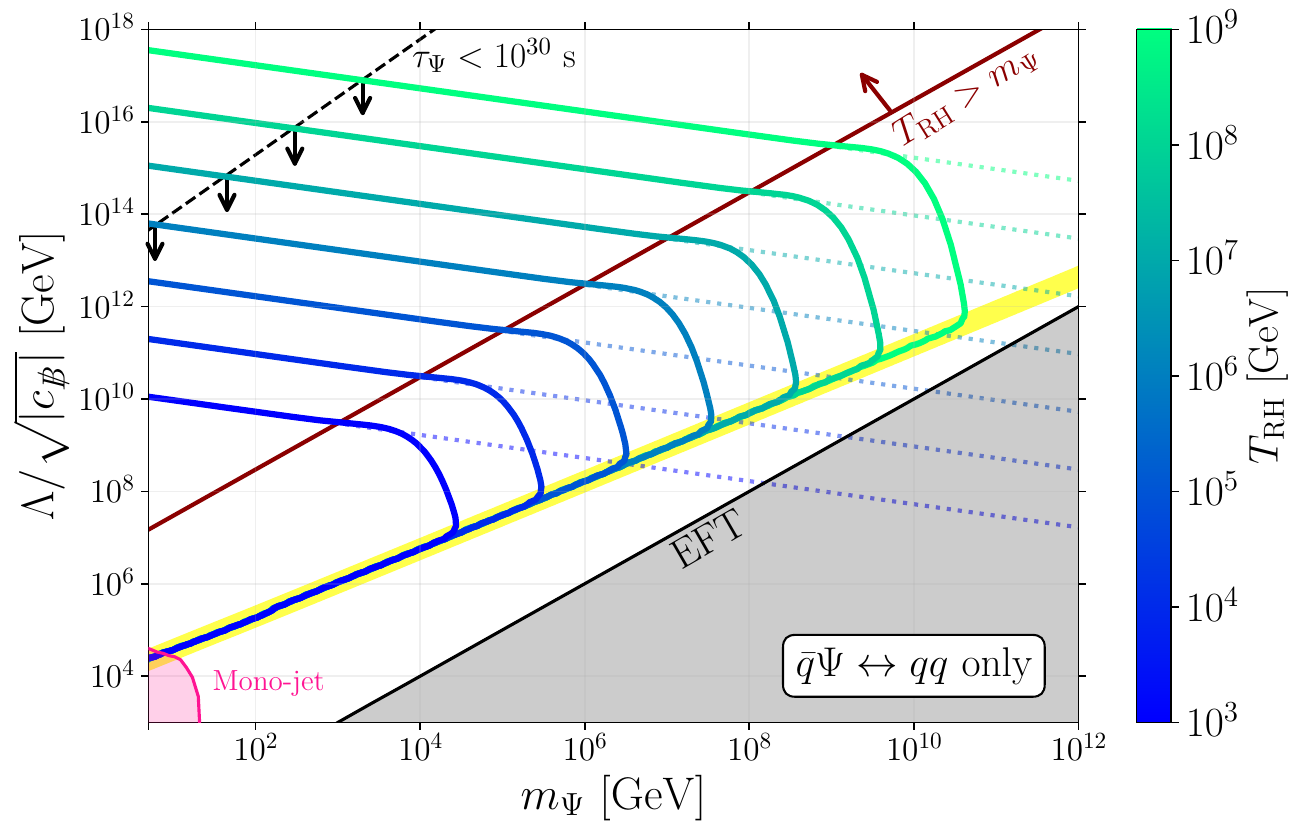}
    \includegraphics[width=0.48\linewidth]{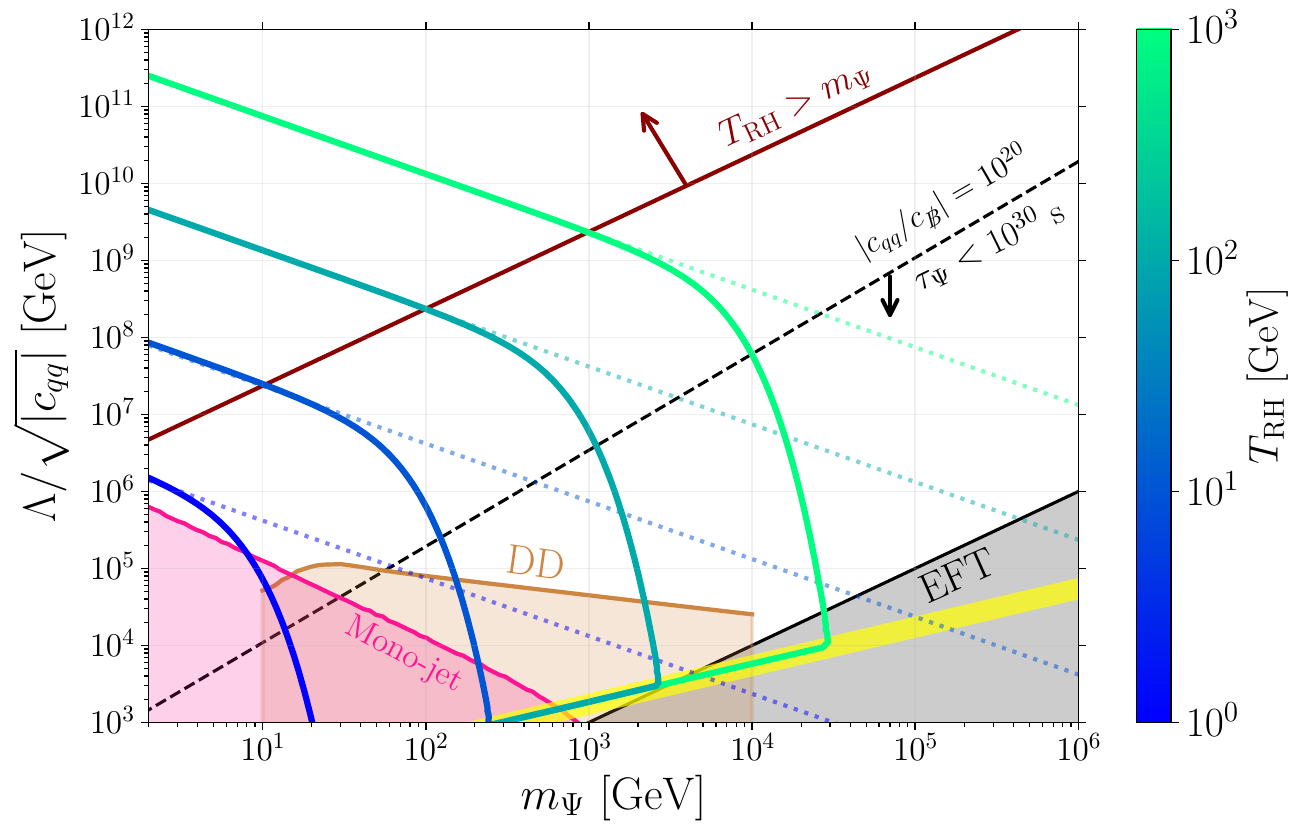}
    \caption{Parameter space reproducing the observed relic abundance, $\Omega h^2 \simeq 0.12$, in the $(m_\Psi,\ \Lambda/\sqrt{|c|})$ plane. \textit{Left panel}: $\bar q \Psi \leftrightarrow qq$ only. \textit{Right panel}: $\Psi\bar\Psi \leftrightarrow q\bar q$–dominated regime. Solid coloured lines show numerical results for different reheating temperatures $T_{\rm RH}$ (colour bar), while dotted lines are analytical freeze-in predictions. The yellow band denotes the freeze-out approximation. The red line marks $T_{\rm RH}=m_\Psi$, separating standard and Boltzmann-suppressed freeze-in. Gray regions are excluded by EFT validity. Dashed black lines indicate $\tau_\Psi=10^{30}$ s. Pink regions show bounds from LHC monojet searches and brown region from DM direct detection.} 
    \label{fig:Oh2}
\end{figure*}

Let us compute the relic density in the pure UV freeze-in limit\,\cite{Elahi:2014fsa} where we assume a negligible initial $n_\Psi$ and we limit to the case $T_\text{RH} \gg m_\Psi$.\footnote{The \textit{freeze-in at stronger coupling} coupling regime (also known as Boltzmann suppressed freeze-in)~\cite{Cosme:2023xpa,Cosme:2024ndc,Costa:2024ugy,Costa:2025gqf,Bernal:2025fcl}, where $m\gtrsim T_\text{RH}$ will be considered later in Section~\ref{sec:num_res} where the numerical results are discussed. See Fig.\,\ref{fig:Oh2}} Therefore, any term that includes $Y_\Psi$ on the right-hand side of the Boltzmann equation can be neglected. In Sec.\,\ref{sec:uv} we will discuss scenarios where the initial abundance, e.g. from gravitational production, is negligible. Then, the freeze-in Boltzmann equations for the two operators read simply as follows:
\be\label{eq:BoltzmannEq_FI}
H S x \frac{\mathrm{d} Y_\Psi}{\mathrm{d}x} \simeq \gamma_{\bar q\Psi \leftrightarrow qq} + \gamma_{\Psi\bar\Psi \leftrightarrow q\bar q},
\ee
where we took $\Delta\simeq  1$ and for simplicity we have also neglected the decay term $\gamma_{\Psi\to qqq}$. Given the UV nature of the process,  the cross-sections can be evaluated in the limit $s \gg m_\Psi^2$ and simplify from Eqs.\,\eqref{eq:sigma_qPsi_qq} and \eqref{eq:sigma_PsiPsi_qq} to 
\al{\label{eq:sigma_FI}
\sigma_{\bar q\Psi\to qq}(s) &\simeq\frac{|\cB|^2 N_f(s)}{72\pi m_\Psi^2 \Lambda^4} s^2,\\
\sigma_{ \Psi \bar\Psi\to  q\bar q}(s)&\simeq\frac{|c_{qq}|^2N_f'(s)}{6912 \pi m_\Psi^4\Lambda^4}s^3.
}
With these approximations, one can directly integrate Eq.\,\eqref{eq:BoltzmannEq_FI} (see Eqs.\,\eqref{eq:approx_reaction_rates} and \eqref{eq:Boltzmann_FI}). Using the correct value of the DM yield today,
\be\label{eq:DMYield_today}
Y_{\rm crit} = 0.12  \frac{\rho_c}{m_\Psi S_0} \simeq 4.4\times 10^{-10}~\text{GeV}/m_\Psi,
\ee
where $\rho_c\simeq 8.098\times 10^{-11}h^{-2}~\text{eV}^4$ is the critical density today, $S_0\simeq 2.2\times 10^{-11}~\text{eV}^3$ is the entropy density today, and using also $g_*(T_\text{RH})=g_{*s}(T_\text{RH})=106.75$, we obtain
\al{\label{eq:analytical_estimate_FI}
\frac{Y^0_{\Psi qqq}}{Y_{\rm crit}}\biggl|_\text{FI} & \simeq 174\left(\frac{N_f}{4}\right)\left(\frac{2~\text{GeV}}{m_\Psi}\right)\left(\frac{10^{12}~\text{GeV}}{\Lambda/\sqrt{|\cB|}}\right)^4\\
&\hspace{1cm}\times \left(\frac{T_\text{RH}}{10^5~\text{GeV}}\right)^5,\\
\frac{Y^0_{ \Psi\Psi qq}}{Y_{\rm crit}}\biggl|_\text{FI} & \simeq 17\left(\frac{N_f'}{4}\right)\left(\frac{|c_{qq}|}{10^{-5}}\right)^2\left(\frac{2~\text{GeV}}{m_\Psi}\right)^3 \\ 
&\hspace{1cm} \times \left(\frac{10^{12}~\text{GeV}}{\Lambda}\right)^4
\left(\frac{T_\text{RH}}{10^5~\text{GeV}}\right)^7.}
Here, $Y^0_{\Psi qqq}|_\text{FI},~Y^0_{\Psi\Psi qq}|_\text{FI}$ are the yields coming from the processes $ qq\to \bar  q\Psi$ and $q\bar q\to \Psi \bar\Psi$, respectively,  evaluated today using the freeze-in approximation. The total DM yield is equal to the sum of both. We see that, because of the different dependence on the center of mass energy in Eq.\,\eqref{eq:sigma_FI}, the yields above scale differently with the mass and reheating temperature.
Notice that we can pull $N_f,N_f'$ out of the integrals, as they are approximately piecewise functions of $s$.

Let us compare the yields in the freeze-in regime to understand which operator dominates and in what region of the parameter space. Taking the ratio of each contribution in Eq.\,\eqref{eq:analytical_estimate_FI}, which is also the ratio of the relic density we obtain
\begin{equation}
    \frac{Y^0_{\Psi qqq}}{Y^0_{\Psi\Psi qq}} \biggl|_\text{FI}\simeq \frac{58}{51} \left|\frac{\cB}{c_{qq}} \right|^2 \left(\frac{m_\Psi}{T_\text{RH}}\right)^2.
\end{equation}
Setting this ratio to unity we obtain 
\begin{equation}\label{eq:cB_over_cqq}
     \left|\frac{\cB}{c_{qq}}\right| \simeq  \left(\frac{T_{\rm RH}}{m_\Psi}\right).
\end{equation}
We can therefore see that the pure UV freeze-in yields are of the same order when $T_{\rm RH} \sim m_\Psi$ and $|\cB|\sim|c_{qq}|$, however, this is also the regime where the freeze-in approximation we used breaks down and we enter the \textit{freeze-in at stronger coupling} regime\,\cite{Cosme:2023xpa,Costa:2024ugy,Cosme:2024ndc,Costa:2025gqf,Arcadi:2024wwg,Bernal:2025fcl}. In the UV freeze-in regime where $T_{\rm RH} \gg m_\Psi$, we need a large hierarchy between the coupling to have comparable yield production from the two processes, i.e. $|\cB| \gg |c_{qq}| $. Therefore, in the standard freeze-in regime we expect $Y^0_{\Psi\Psi qq}$ to always dominate if the Wilson coefficients are of the same order of magnitude.

\begin{figure}
    \centering
    \includegraphics[width=1\linewidth]{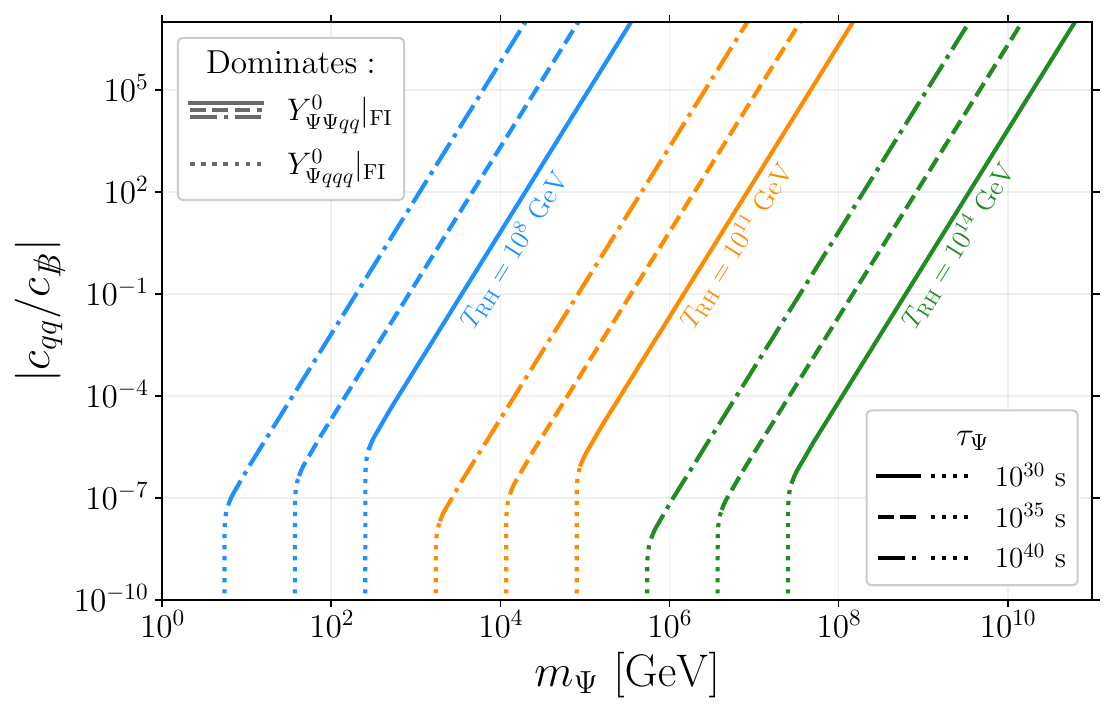}
    \caption{Relation between the different contributions to the DM abundance in the freeze-in approximation from Eq.\,\eqref{eq:analytical_estimate_FI}. The lines denote where the DM abundance is reproduced and the lifetime is fixed to $\tau_\Psi=10^{30}$s (solid+dotted), $10^{35}$s (dashed+dotted) and $10^{40}$s (dashed-dotted+dotted). For solid, dashed and dashed-dotted the abundance is dominated by $Y^0_{\Psi \Psi qq}|_\text{FI}$, whereas for dotted it is dominated by $Y^0_{\Psi qqq}|_\text{FI}$. We show these results for $T_\text{RH}=10^{8}$ GeV (blue), $10^{11}$ GeV (orange) and $10^{14}$ GeV (green). The regions left to each curve are overabundant.}
    \label{fig:YFI_Analytical}
\end{figure}

In Fig.\,\ref{fig:YFI_Analytical} we show the ratio between the couplings $|c_{qq}/\cB|$ as a function of the mass based on our results for $Y^0_{\Psi qqq}|_\text{FI}$ and $Y^0_{\Psi\Psi qq}|_\text{FI}$. Each line corresponds to $Y^0_{\Psi qqq}|_\text{FI}+Y^0_{\Psi\Psi qq}|_\text{FI}=Y_\text{crit}$, that is, the total DM abundance is being reproduced, whereas colours denote different choices of reheating temperature. The solid, dashed and dashed-dotted curves indicate regimes in which $Y^0_{\Psi \Psi qq}|_\text{FI}$ dominates the DM production for different choices of the DM lifetime $\tau_\Psi$. The lines turn to dotted instead once $Y^0_{\Psi qqq}|_\text{FI}$ dominates. In this latter case, we see that the lines become vertical, as there is a unique value of $|\cB|$ that satisfy $Y^0_{\Psi qqq}|_\text{FI}=Y_\text{crit}$ for a fixed lifetime. As stressed, the process $\bar q \Psi \leftrightarrow qq$ dominates for small values of $|c_{qq}/\cB|$, as it is clear from the figure.

\subsection{Freeze-out approximation}

We can now study the opposite regime where the DM interacts strongly enough to be in thermal equilibrium with the SM plasma before freezing-out. As in the freeze-in case, we will discuss the analytical approximation for both processes $\bar q \Psi \leftrightarrow qq$ and $\Psi\bar\Psi\leftrightarrow q\bar q$ and compute the DM abundance. The standard freeze-out condition is defined as the temperature $x_{\rm FO}$ at which the interaction rate is comparable to the Hubble parameter:
\begin{align}\label{eq:xFO_def}
    \Gamma(x_{\rm FO}) &\equiv  n^{\rm eq} (x_{\rm FO})\langle \sigma v \rangle(x_{\rm FO}) = H(x_{\rm FO}). 
\end{align}
This is a good estimate for the departure from equilibrium for a 'fast' freeze-out, i.e. almost instantaneous, as it is the case for $\Psi \bar\Psi \leftrightarrow  q\bar q$. When the freeze-out is much slower instead, for instance for $\bar q\Psi \leftrightarrow q q$, as we will see shortly, a better approximation for the departure from thermal equilibrium is given by the decoupling time $x_\text{dec}$ defined as\,\cite{Kramer:2020sbb}
\footnote{The process $\bar q\Psi \leftrightarrow qq $ resembles the \textit{zombie DM} scenario considered in Ref.\,\cite{Kramer:2020sbb}. The crucial difference is that here the bath particle producing the DM is fully relativistic.}
\begin{align}\label{eq:xdec_def}
    \Gamma(x_{\rm dec}) &\equiv  n^{\rm eq} (x_{\rm dec})\langle \sigma v \rangle(x_{\rm dec}) = H(x_{\rm dec}) x_{\rm dec}.
\end{align}
The difference can be highlighted by comparing the interaction rates for both processes, as derived carefully in App.\,\ref{app:boltzmannEq},
\al{\label{eq:FO_rates_approx}
\Gamma_{\Psi qqq} & = n^\text{eq}_q \langle\sigma_{\bar q\Psi\rightarrow qq}v\rangle\simeq \frac{8N_f(m_\Psi) |c_{\slashed{B}}|^2m_\Psi^5}{e^2\pi^3\Lambda^4}\frac{1}{x^3},\\
\Gamma_{\Psi\Psi qq} & = n^\text{eq}_\Psi\langle \sigma_{\Psi\bar\Psi\rightarrow q\bar q}v\rangle\\
& \simeq \frac{85 \sqrt{17}e^{-17/16}}{73728\sqrt{2}\pi^3}\frac{|c_{qq}|^2N_f'(m_\Psi)^2m_\Psi^5e^{-x}}{x^{3/2}\Lambda^4}.
}
In fact, since $\Gamma_{\Psi qqq}$
decreases polynomially while $\Gamma_{\Psi\Psi qq}$
decreases exponentially with $x$, the former freeze-out is much slower, justifying the use of Eq.\,\eqref{eq:xdec_def} to characterise the departure from thermal equilibrium.
Using Eqs.\,\eqref{eq:xFO_def}-\eqref{eq:FO_rates_approx}, we can compute $x_\text{dec}$ and $x_\text{FO}$:
\al{
    \label{eq:x_FO_estimate}
x_\text{dec}^2 & \simeq \frac{24N_f(m_\Psi)|c_{\slashed{B}}|^2}{e^2\pi^3}\sqrt{\frac{10}{g_*(m_\Psi/x_\text{dec})}}\frac{m_\Psi^3M_P}{\Lambda^4},\\
x_\text{FO} & \simeq \log \left(\frac{85^{3/2}e^{-17/16}|c_{qq}|^2{N_f'(m_\Psi)}^2M_Pm_\Psi^3}{24576\pi^4\sqrt{g_*(m_\Psi/x_\text{FO})}\Lambda^4}\right).
}
At late times, i.e. $x \gg x_{\rm FO} \; (x_{\rm dec})$, when $\Gamma \ll H$ ($\Gamma \ll x H$), we have that $Y_{\Psi} \gg Y_{\Psi}^{\rm eq}$ and therefore we can solve the simplified Boltzmann equation\,\cite{Kolb:1990vq}
\al{
\label{eq:YFO_estimate}
\frac{\dd Y}{\dd x} \simeq &  \frac{\Gamma_{\Psi\Psi qq}}{x H Y^{\rm eq}_{\Psi}} Y^2 + 
\frac{\Gamma_{\Psi qqq}}{x H} Y.
}
By considering one dominating term at a time, we can solve the equation above by separation of variables, leading to the corresponding yields today (see Eqs.\,\eqref{eq:Yqq_FO_apprx} and \eqref{eq:Yqqq_FO_approx}) 
\begin{align}\label{eq:analytical_estimate_FO}
& \frac{Y^0_{\Psi\Psi qq}}{Y_{\rm crit}}\Bigg|_\text{FO}  \simeq 1.2 \times 10^{13} ~\text{GeV}^{-1} \frac{\sqrt{g_*} x_\text{FO}\Lambda^4}{|c_{qq}|^2g_{*s} M_P m_\Psi^2 {N_f'}^2} \nonumber\\
&\hspace{1.5cm} \simeq 1.3 \left(\frac{4}{|c_{qq}|N_f'}\right)^2\left(\frac{x_\text{FO}}{20}\right)\left(\frac{g_*}{50}\right)^{1/2}\left(\frac{50}{g_{*s}}\right) \nonumber \\
&\hspace{2.5cm}\times \left(\frac{\Lambda}{500~\text{GeV}}\right)^4\left(\frac{200~\text{GeV}}{m_\Psi}\right)^2,\\
&\frac{Y^0_{\Psi q qq}}{Y_{\rm crit}}\Bigg|_\text{FO}  \simeq 1.3\times 10^9~\text{GeV}^{-1}\frac{m_\Psi x_\text{dec}^{3/2}e^{-2x_\text{dec}}}{g_{*s}}, \nonumber\\
&\hspace{0.8cm} \simeq 1.0 \left(\frac{m_\Psi}{50~\rm GeV}\right) \left(\frac{x_\text{dec}}{12}\right)^{3/2}e^{-2 \left(x_\text{dec}-12\right)}\left(\frac{100}{g_{*s}}\right), \nonumber
\end{align}
where we again emphasise the very distinct dependence on the parameters due to the different freeze-out dynamic of the processes. In particular, $Y^0_{\Psi q qq}|_\text{FO}$ depends exponentially on $x_\text{dec}$, making it very sensitive to changes in the parameters around the freeze-out region. Using the expressions above, we find out that the contribution from $Y_{\Psi qqq}^0|_\text{FO}$ tends to be larger than $Y_{\Psi \Psi qq}^0|_\text{FO}$ for larger masses, but becomes completely negligible in the mass range of interest once $|c_{qq}/\cB|\gtrsim 10^{15}$.\\

\subsection{Numerical results}
\label{sec:num_res}
Having gained an analytical understanding for the DM production in this model, we can now discuss the numerical results and compare them with the analytical approximations. We numerically solve the Boltzmann Eq.\,\eqref{eq:Boltzmann_eq_general} neglecting the decay term, while fully accounting for the energy dependence of $N_f,N_f'$ and also keeping track of $\Delta$.

The results are shown in Fig.\,\ref{fig:Oh2}, where we show the parameter space reproducing the observed DM relic abundance. In the left panel we consider only the process $\bar q \Psi\leftrightarrow qq$, whereas in the right panel we consider both but in the regime in which $\Psi\bar\Psi \leftrightarrow q\bar q$ dominates (see Fig.\,\ref{fig:YFI_Analytical} and Eq.\,\eqref{eq:cB_over_cqq}). As in each case one operator dominates the DM production, we show the plot as a function of the cut-off $\Lambda$ normalized by the corresponding coupling. The coloured solid lines represent the values of masses and couplings that reproduce the correct relic abundance, i.e. $\Omega h^2 \simeq 0.12$, for different values of the reheating temperature $T_{\rm RH}$. The dotted lines of the same colour are the analytical freeze-in result and the yellow lines are the freeze-out analytical approximations. We see that analytical results describe well the full numerical result in their range of validity. Let us underline the presence of a regime of \textit{freeze-in at stronger coupling} (also known as Boltzmann suppressed freeze-in)\,\cite{Cosme:2023xpa,Costa:2024ugy,Cosme:2024ndc,Costa:2025gqf,Arcadi:2024wwg,Bernal:2025fcl}. This regime connects the standard freeze-in and freeze-out regimes and it is characterised by $m_{\Psi}>T_{\rm RH}$, where the number density of the bath particles producing the DM is Boltzmann suppressed. Therefore, the yield is exponentially suppressed and the UV scale can be much smaller compared to the UV freeze-in case. Notice that the freeze-out is a lower limit for $\Lambda$ since, once the UV scale reaches those values, $\Psi$ thermalises and all freeze-in lines converge onto the freeze-out curve. In Fig.\,\ref{fig:Oh2} we denote by the red lines approximately the point of transition between the freeze-in and the Boltzmann suppressed regime, i.e. $T_{\rm RH} = m_{\Psi}$.

The gray areas, denoted as ``EFT", are regions of the parameter space theoretically excluded because the EFT validity condition $m_{\Psi}<\Lambda/|c_{qq,\slashed{B}}|^{1/2}$ is violated. Let us also notice that the EFT remains valid throughout the freeze-in parameter space in both cases, since $\Lambda/|c_{qq,\slashed{B}}|^{1/2} > T_{\rm RH}$ is satisfied everywhere in this region. In the case of freeze-out via $\Psi \bar{\Psi} \leftrightarrow q \bar{q}$ (right panel), the relevant parameter space is excluded by EFT validity for $m_\Psi \gtrsim 2~\mathrm{TeV}$\,\cite{Griest:1989wd}. By contrast, for $\bar{q} \Psi \leftrightarrow qq$ only (left panel), the entire freeze-out line lies within the EFT regime of validity\,\cite{Kim:2019udq,Kramer:2020sbb}.

In Fig.\,\ref{fig:Oh2} we also include the relevant constraints to our model, whereas we defer a detailed discussion of their derivation to the following section. 
The dashed black lines indicate a lifetime of $\tau_\Psi = 10^{30}\,\mathrm{s}$; regions below this line are constrained by indirect detection. In particular, in the left panel this line is fixed, as both production and decay are controlled by $\Lambda/\sqrt{|\cB|}$, while in the right panel its depends on the ratio $|c_{qq}/\cB|$. As a consequence, we see that most of the parameter space of the process $\bar q\Psi\leftrightarrow qq$ is excluded by indirect detection if $\Psi$ constitutes the total DM abundance.  By contrast, for $\Psi\bar\Psi\leftrightarrow q\bar q$ we can tune $\cB$ to make $\Psi$ long-lived enough. 
Then, the pink shaded areas denote bounds from monojet searches at the LHC and the brown area the bound from direct detection experiments. Both are relevant for low values of the cut-off and masses, and are complementary to the EFT bound.\\

%%%%%%%%%%%%%%%%%%%%%%%%%%%%
%%%%%%%%%%% Pheno %%%%%%%%%%
%%%%%%%%%%%%%%%%%%%%%%%%%%%%

\section{Phenomenology}\label{sec:pheno}

In this section we analyse the phenomenological implications of the effective interactions introduced above. Our goal is to determine whether the spin-3/2 particle can account for the observed DM abundance while remaining consistent with current experimental constraints. Therefore we confront the viable parameter space with experimental searches, including indirect detection, direct detection, and collider probes. In particular, we examine the interplay between the BNV operator in Eq.\,\eqref{eq:La_dim6_BNV} and the baryon number conserving operators in Eq.\,\eqref{eq:La_dim6_qq}, highlighting how they control respectively the DM decay properties and scattering signatures.

This combined analysis allows us to identify the regions of parameter space, shown in Fig.\,\ref{fig:Oh2}, where the correct relic abundance is achieved and to assess the complementarity of different experimental probes in testing the model.

\subsection{Indirect detection}

Indirect detection is a way of probing DM through the observation of SM particles produced in its decay or annihilation in astrophysical environments\,\cite{Hooper:2018kfv}. In our scenario, the relevant signal arises from DM decaying via the BNV operator into stable SM final states such as photons, neutrinos, positrons, and anti-protons, whose spectra can be compared with existing experimental bounds. Since the propagation of charged particles is more involved, we focus the discussion on experiments measuring gamma-rays and neutrinos.

\begin{figure}
    \centering
    \includegraphics[width=\linewidth]{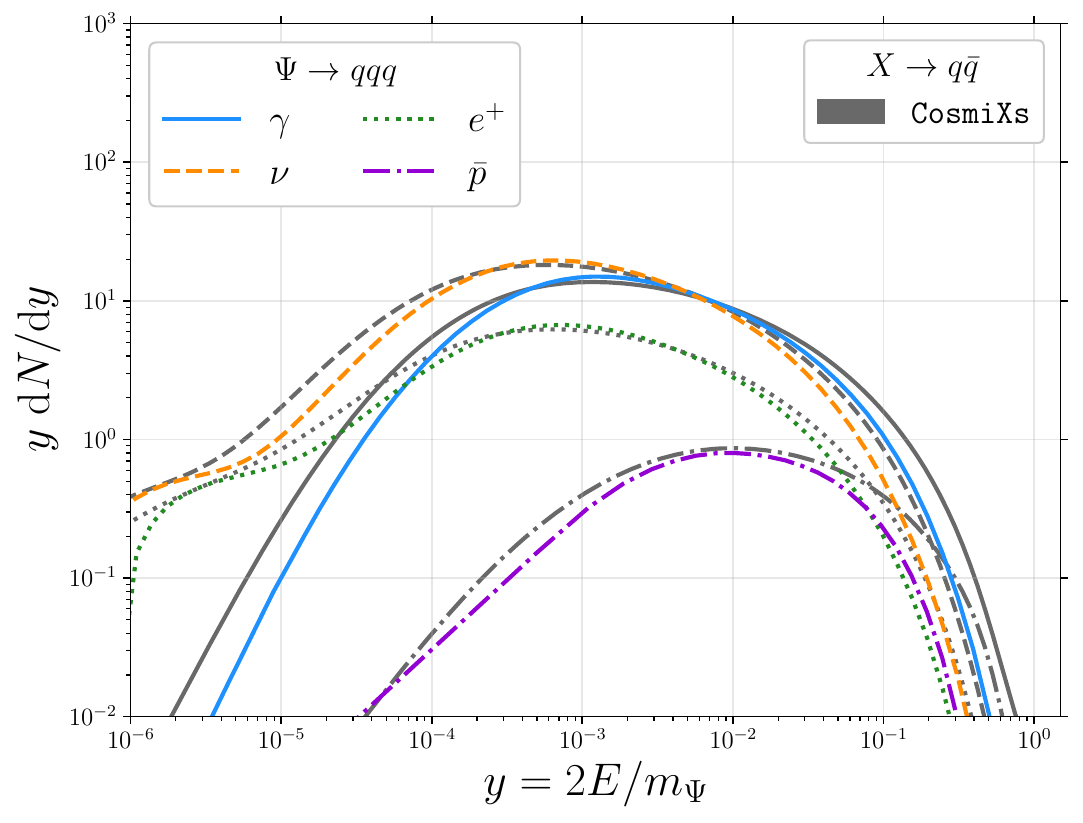}
    \caption{Differential spectra at production for the decaying spin-3/2 particle of mass $m_\Psi=1$ TeV as a function of the energy fraction $y=2E/m_\Psi$. We show the spectrum for photons (solid blue), neutrinos (dashed orange), positrons (dotted green) and anti-protons (dot-dashed purple). The corresponding gray lines are the ones obtained from \texttt{CosmiXs}\,\cite{Arina:2023eic} for a decaying particle $X\to q\bar q$.} 
    \label{fig:Spectrum}
\end{figure}

To accurately determine the expected signal, we first compute the full decay spectrum of the DM candidate using Monte Carlo simulations. We implement our EFT operators in \texttt{FeynRules}\,\cite{Alloul:2013bka,Degrande:2011ua,Christensen:2008py,Christensen:2013aua} and simulate the underlying partonic events via \texttt{MadGraph5\_aMC}\,\cite{Alwall:2014hca,Alwall:2011uj}, which is subsequently interfaced with \texttt{Pythia8}\,\cite{Bierlich:2022pfr} for parton showering, hadronisation and decay of unstable SM particles. This procedure provides the differential energy spectra $\dd N/\dd E$, shown in Fig.\,\ref{fig:Spectrum}, for all relevant final states taking into account the full cascade of secondary decays.

We then compare the resulting spectra with those obtained in benchmark scenarios commonly used in the literature. In particular, we consider the case of a DM particle decaying into quark--antiquark pairs $q\bar{q}$, which is a standard reference channel in indirect detection studies\,\cite{Cirelli:2010xx,Arina:2023eic}. We use \texttt{CosmiXs}\,\cite{Arina:2023eic} to generate the spectrum for the $q\bar{q}$ case. As shown for example in Fig.\,\ref{fig:Spectrum} for $m_\Psi=1$ TeV, the spectra obtained in our model closely resemble those of the $q\bar{q}$ channel over the relevant energy range. This similarity can be understood from the fact that, after hadronisation, different quark-initiated final states tend to produce universal spectra dominated by pion production and subsequent decays into photons and leptons.

Given this spectral degeneracy, we can reliably reinterpret existing bounds derived for the $q\bar{q}$ channel. Current indirect detection searches, in particular from gamma-ray of diffuse emission constrain decaying DM scenarios to lifetimes of order
\begin{equation}\label{eq:ID_constraint}
    \tau_{\rm DM} \gtrsim 10^{30}\ \text{s} \, ,
\end{equation}
which approximately saturates the current gamma-ray constraints\,\cite{Blanco:2018esa,IceCube:2022clp,Chianese:2021jke,Aloisio:2025nts,Boehm:2025qro,Bhattacharya:2019ucd,Ishiwata:2019aet} in a wide mass range. We note that adopting this bound is a conservative choice since it can overconstrain the parameter space; however, this choice ensures that we do not underestimate the impact of indirect detection bounds. Since our predicted spectra do not significantly deviate from the benchmark case, we adopt this bound as a good approximation for our model and it is presented in Fig.\,\ref{fig:Oh2} for the two different operators.

As stressed before, the indirect detection bound is particularly strong for the scenario where DM is generated solely from $\bar q\Psi\leftrightarrow qq$, since both production mechanism and decay are controlled by the same coupling. As it is evident from the left panel of Fig.\,\ref{fig:Oh2}, setting the bound in Eq.\,\eqref{eq:ID_constraint} excludes all the region below the black dashed line, in particular excluding the whole freeze-out, Boltzmann suppressed freeze-in and a large portion of the UV freeze-in regions. It should be noted that if the spin-3/2 particle does not constitute the entirety of the DM abundance, for instance, in regions of parameter space below the freeze-out line, the indirect detection bound is correspondingly weakened. For the regime where the abundance is controlled by $\Psi\bar\Psi \leftrightarrow q\bar q$ the bound depends crucially on the ratio $|c_{qq}/\cB|$. In Fig.\,\ref{fig:Oh2} we show for example the case when $|c_{qq}/\cB|=10^{20}$, for which indirect detection still allows for freeze-in and part of the Boltzmann suppressed freeze-in.

\subsection{Direct detection}
The direct detection experiments measure the recoil energy deposited by DM particles from the Galactic halo scattering off nuclei $N$. These searches are most sensitive at low cut-off scales, i.e. near the freeze-out regime. In the scenario where $\bar q \Psi \leftrightarrow qq$ controls dark matter production, direct detection constraints are therefore irrelevant: the entire freeze-out region is already excluded by indirect detection bounds. We thus omit direct detection limits for this operator. The situation is qualitatively different for the process \(  \Psi\bar \Psi \leftrightarrow  q\bar  q \). In this case, the indirect detection constraints become weaker at small values of \(c_{\slashed{B}}\), while direct detection is controlled by the independent Wilson coefficient \(c_{qq}\).

The spin-independent DM--nucleus scattering cross-section at zero momentum transfer is given by\,\cite{Ding:2013nvx}
\begin{equation}
    \sigma^{\rm SI}_{\Psi N} = \frac{f_N^2 c_{qq}^2}{\pi \Lambda^4} \frac{m_{\Psi}^2 m_N^2}{(m_{\Psi}+m_N)^2},
\end{equation}
with $m_N$ the mass of the nucleon and $f_N$ is the effective coupling that takes into account the contributions of valence quarks of nucleons
\al{
& f_N = Z b_p + (A - Z) b_n, \\
& b_p = b_n = 3.
}
The region excluded by current direct detection data, derived from the LZ~2022 results\,\cite{LZ:2022lsv}, is shown as the brown shaded area in Fig.\,\ref{fig:Oh2}. These bounds, in complementarity with monojet searches, exclude the entire low-mass freeze-out region. At larger DM masses, the parameter space is instead constrained by the validity of the effective field theory, which ultimately excludes the freeze-out regime for this operator.\footnote{An independent way of excluding the freeze-out for spin-3/2 particles was recently obtained in Ref.\,\cite{Goldstein:2026iuu}, where instead cosmological observables are used.}

Direct detection experiments are also already sensitive to parts of the \textit{freeze-in at stronger coupling} parameter space in the mass range \(10 \,\mathrm{GeV} \lesssim m_{\Psi} \lesssim 10^4 \,\mathrm{GeV}\), probing effective scales up to \(\Lambda/\sqrt{c_{qq}} \sim 10^5 \,\mathrm{GeV}\). Future experiments such as DARWIN\,\cite{DARWIN:2016hyl} are expected to extend this reach by approximately one order of magnitude in the cut-off scale. Overall, direct detection provides a powerful probe that already tests and excludes significant portions of parameter space in both the freeze-out and freeze-in regimes.

\subsection{Monojet searches}

% \GMS{forgot to mention these bounds are independent if they are DM or not}

We now turn to collider searches, focusing in particular on monojet signals, which consist of a single jet plus missing energy in the final state. In our model, both the BNV operator in Eq.\,\eqref{eq:La_dim6_BNV} as well as the operators in Eq.\,\eqref{eq:La_dim6_qq} can generate such signature at the LHC, although in different ways. For the latter case, the relevant process is the annihilation of the two initial partons in two DM particles plus a jet radiated from the initial state, i.e. $pp\to \Psi\bar \Psi+\text{jet}$, which is a five-body scattering. This signature in the spin-3/2 scenario was previously studied in Ref.\,\cite{Ding:2013nvx} using Run 1 data. For the former case instead, as the interaction involves only one DM and three quarks, one can obtain the monojet signature directly from a 2-to-2 scattering, $pp\to \Psi +\text{jet}$. This type of channel was explored in some specific models for spin-1/2 particles\,\cite{Dutta:2014kia, Allahverdi:2017edd,Bittar:2024fau,Bittar:2024nrn}, nonetheless mostly presented in terms of specific UV completion. In what follows we will revisit these bounds and obtain the corresponding constraints to the parameter space.

The CMS and ATLAS collaborations have searched for monojet signals in Ref.\,\cite{ATLAS:2021kxv,CMS:2021far} using Run 2 data. Using the data provided in these references, we see that the uncertainty in the number of events is around $\mathcal{O}({1\%})$, with the total number of events being roughly $N_\text{tot}^\text{LHC}\sim \mathcal{O}(10^6)$. We can therefore use the cross-sections in Eqs.\,\eqref{eq:sigma_qPsi_qq} and \eqref{eq:sigma_PsiPsi_qq} to estimate the sensitivity to the model parameters. Considering only the high-energy behaviour of the cross-sections $s\gg m_\Psi^2$ and using a luminosity of $\mathcal{L}=139~\text{fb}^{-1}$, we obtain
\begin{align}
\label{eq:monojet_estimate}
&\frac{\mathcal{L}\sigma_{\bar q\Psi\to qq}}{N_\text{tot}^\text{LHC}}
\sim  \; 10^{-2} \left(\frac{\sqrt{s}}{1~\text{TeV}}\right)^4\left(\frac{1~\text{GeV}}{m_\Psi}\right)^2\left(\frac{10^2~\text{TeV}}{\Lambda/\sqrt{|\cB|}}\right)^4, \nonumber\\
&\frac{g_s^2}{4\pi}\frac{\mathcal{L}\sigma_{\Psi \bar \Psi\to qq}}{N_\text{tot}^\text{LHC}}
\sim \; 10^{-2}\left(\frac{\sqrt{s}}{1~\text{TeV}}\right)^6 \\
& \qquad \qquad \qquad \times  \; \left(\frac{1~\text{GeV}}{m_\Psi}\right)^4\left(\frac{5\times 10^2~\text{TeV}}{\Lambda/\sqrt{|c_{qq}|}}\right)^4, \nonumber
\end{align}
where for $\sigma_{\Psi \bar \Psi\to qq}$ we added a factor of $g_s^2/4\pi$, with $g_s$ the strong coupling, to take into account the emission of the jet from the initial quarks. From the estimates above, we see that, even though the process $pp\to \Psi\bar \Psi+\text{jet}$ has an extra phase-space suppression, it is still expected to constraint larger values of the cut-off than $pp\to \Psi +\text{jet}$ for $m_\Psi\sim 1$ GeV and $\cB\sim c_{qq}$. This is nothing but a consequence from the fact that $\sigma_{\Psi \bar \Psi\to qq}/\sigma_{\bar q\Psi\to qq}\propto m_\Psi^{-2}$, thus growing faster for lower masses. For larger masses one would consequently expect $pp\to \Psi +\text{jet}$ become more relevant, however one must take into consideration that the same operator can induce the decay of the spin-3/2. With Eq.\,\eqref{eq:life-time} we can estimate the laboratory decay length as
\be
\! \! \! \gamma c \tau_\Psi \sim 10^8\text{m}\left(\frac{E_\Psi}{1~\text{TeV}}\right)\left(\frac{1~\text{GeV}}{m_\Psi}\right)^6\left(\frac{\Lambda/\sqrt{|\cB|}}{10^2~\text{TeV}}\right)^4,
\ee
where $E_\Psi$ is the typical energy of the spin-3/2 and $\gamma=E_\Psi/m_\Psi$ the corresponding boost factor. From the estimate above and taking into account the size of the ATLAS and CMS detectors of $\mathcal{O}(10~\text{m})$, we see that we expect to sharply lose sensitivity for masses above $m_\Psi\gtrsim 10$ GeV.\footnote{One way to extend the sensitivity would be to consider in addition displaced jet signatures, which would, however, demand a dedicated analysis.}

In order to have a more precise evaluation of this bound, we simulate events in \texttt{MadGraph5\_aMC}\,\cite{Alwall:2014hca,Alwall:2011uj} at parton level implementing the cuts of Refs.\,\cite{ATLAS:2021kxv,CMS:2021far} and perform a $\chi^2$ test using the observed distribution of the number of events in bins of missing transverse energy. The 90\%C.L. exclusion regions are shown in Fig.\,\ref{fig:Oh2} in pink. In the right panel, for $pp\to \Psi\bar \Psi+\text{jet}$, we see that the bound on $\Lambda/\sqrt{|c_{qq}|}$ is strongest at $m_\Psi\sim 1$ GeV, afterwards decreasing until $m_\Psi\sim 1$ TeV, after which there is not enough energy to produce the spin-3/2 anymore. For the left panel instead, as argued above, we lose sensitivity around $m_\Psi\sim 10$ GeV because the spin-3/2 is not long-lived enough. To obtain this behaviour, we have adopted a simplified strategy of penalizing the predicted number of events by the average probability of the particle decaying outside of the detector, with a typical detector size of $10$ m. In both cases our estimates in Eq.\,\eqref{eq:monojet_estimate} are in very good agreement with the numerical analysis.

%%%%%%%%%%%%%%%%%%%%%%%%%%%%
%%%%% UV completion %%%%%%%%%%
%%%%%%%%%%%%%%%%%%

\section{Toy UV completion}\label{sec:uv}

We now present a toy UV model that can generate the BNV interaction in Eq.\,\eqref{eq:La_dim6_BNV} and also the baryon number conserving ones in Eq.\,\eqref{eq:La_dim6_qq}, as well as avoid catastrophic production in the early universe, while maintaining the consistency of the EFT.

The toy UV model we propose is that of a dark strongly interacting sector, from which the spin-3/2 will emerge as the lightest, stable, composite state. To achieve that, we simply need a new $SU(3)_\text{dc}$ gauge group with one flavour of fundamental non-chiral spin-1/2 particles $\psi$. Due to the similarity with QCD, we refer to the new quantum numbers as \textit{dark colour} and the $\psi$ as \textit{dark quark}. It is then easy to understand that the lightest dark baryon is indeed a spin-3/2 state: given that the state $\psi\psi\psi$ must have the dark colour contracted fully anti-symmetrically via the Levi-Civita tensor, the spin wave-function must be totally symmetric, which carries spin 3/2. The spectrum of QCD with a single flavour was previously studied for instance in Refs.\,\cite{Creutz:2006ts,Farchioni:2007dw}, whose conclusion is that the spin-3/2 baryon $\Psi_\mu$ has a typical mass of $m_\Psi\sim 10\Lambda_\text{QCD}$, where $\Lambda_\text{QCD}$ is the QCD confinement scale. Moreover, there are no light pions in this theory, only the $\eta'$. For our purposes, we assume the same relation holds in the case of the dark QCD scenario we consider, that is $m_\Psi\sim 10\Lambda_\text{dQCD}$.

A remark is in order. With this composite scenario we can avoid the uncontrolled creation of spin-3/2 particles in the early universe\,\cite{Kolb:2021xfn}: We simply require the scale of dark confinement to be lower than the one of inflation and that confinement takes place before reheating ends, such to keep the results of Sec.\,\ref{sec:DM} valid. Unavoidably, the dark quarks will be produced instead, whose initial abundance could contribute to the evolution of the spin-3/2 abundance after confinement. Since fermion production is typically suppressed for masses below the inflationary scale, we neglect this contribution for simplicity.

Having set the framework of the dark QCD model, we can now add further interactions in this model in order to generate the operator in Eq.\,\eqref{eq:La_dim6_BNV}. First, assuming $\Lambda_\text{dQCD}>\Lambda_\text{QCD}$, we can consider the BNV operator for energies $E$ larger than the dark confinement scale, for which we expect the composite structure to break down:
\be
\frac{1}{\Lambda^2}(\bar \Psi_\mu Q)(\bar Q^c \gamma^\mu d_R)\xrightarrow{E\gg \Lambda_\text{dQCD}} \frac{1}{\tilde \Lambda^5} \bar Q^c Qd_R \bar \psi\bar\psi\bar\psi,
\ee
with $\Lambda^2\sim \tilde \Lambda^5/\Lambda_\text{dQCD}^3\sim 10^3\tilde\Lambda^5/m_\Psi^3$. Therefore, we can build a model that generates directly the dimension nine  operator above, which, after dark confinement, is expected to produce the BNV operator. Let us introduce two types of scalars $\Phi_K$ and $\Phi'$, where for the first we can have $K=1,\cdots,N_\Phi$ copies, whose quantum numbers are
\be
\Phi_K\sim (\bar{\bm 3},\bar{\bm 3},\bm 2, -1/6),\quad \Phi'\sim (\bar{\bm 3}, \bar{\bm 3}, \bm 1, 1/3),
\ee
under $SU(3)_\text{dc}\times SU(3)_c\times SU(2)_L\times U(1)_Y$. We can build the following interactions:
\al{\label{eq:LUV}
\La_\text{dQCD} &\supset y_K^i\bar Q^c_i \Phi_K \psi + {y'}^j \bar d_{Rj}^c \Phi' \psi + h.c.\\
&\hspace{1cm}+\mu_{IJ}\Phi_I\Phi_J\Phi',
}
where $y_K^i,{y'}^j$ are flavour-dependent couplings and $\mu_{IJ}$ is a dimensionful coupling, all assumed real for simplicity. Notice that $\mu_{IJ}$ is antisymmetric in $IJ$ due to the contraction of $\Phi_I\Phi_J$ with a $SU(2)_L$ Levi-Civita. Integrating out the scalars, assumed degenerate for simplicity, we can generate the desired dimension nine operator:
\al{\label{eq:dim9_op}
&\La_\text{dQCD} \\ \to & \La_\text{EFT}\supset \frac{\mu_{IJ}}{m_\Phi^6}y_I^iy_J^j{y'}^k(\bar Q^c_i \psi)(\bar Q^c_j \psi)(\bar d_{Rk}^c \psi)+h.c.,
}
where $m_\Phi$ is the mass of the scalars and we can then associate $\tilde \Lambda^5\sim m_\Phi^6/\mu$. Due to the antisymmetry of $\mu_{IJ}$, the overall Wilson coefficient of Eq.\,\eqref{eq:dim9_op} is antisymmetric in the $\bar Q^c Q$ flavours.

Besides the operator in Eq.\,\eqref{eq:dim9_op}, we expect several others at lower dimension, in particular at dimension six. Using \texttt{Matchete}\,\cite{Fuentes-Martin:2022jrf}, we can integrate out the scalar at tree-level. The only operators we get at the level of dimension six are
\be
\La_\text{EFT}^{(6)}=\frac{y_I^iy_I^j}{m_\Phi^2}(\bar Q^c_i \psi) (\bar \psi Q^c_j) + \frac{{y'}^k{y'}^l}{m_\Phi^2}(\bar d_{Rk}^c \psi) (\bar \psi d_{Rl}^c).
\ee
After the confinement from the dark colour gauge group, we can expect the dark quark bilinears to be substituted as $(\bar q^c\psi)( \bar \psi q^c ) \to (\bar q^c\Psi_\mu) (\bar \Psi^\mu q^c)$, which would then be equivalent to the Lagrangian presented in Eq.\,\eqref{eq:La_dim6_qq} with the correspondance $c_{qq}/\Lambda^2\sim y^2/m_\Phi^2$.\footnote{We naturally expect other possible operators not shown in Eq.\,\eqref{eq:La_dim6_qq} to be generated during dark confinement as well.} Our estimates for the Wilson coefficients based on the UV model parameters thus read
\begin{align}
\frac{|c_{qq}|}{\Lambda^2} & \sim \frac{1}{(10^{12}~\text{GeV})^2} \left(\frac{y}{1.0}\right)^2 \left(\frac{10^{12}~\text{GeV}}{m_\Phi}\right)^2,\\ \nonumber
\frac{|c_{\slashed{B}}|}{\Lambda^2} & \sim \frac{1}{(10^{31}~\text{GeV})^2} \left(\frac{y}{1.0}\right)^3 \left(\frac{m_\Psi}{1~\text{GeV}}\right)^3\left(\frac{(10^{12}~\text{GeV})^5}{m_\Phi^6/\mu}\right).
\end{align}
Undoubtedly, the precision of our estimates is limited by our lack of knowledge of the $SU(3)_\text{dc}$ confinement dynamics. Notwithstanding, we can still have a general idea of how the EFT coefficients scale with the UV parameters. In particular, we see that $\cB$ is expected to be much more suppressed than $c_{qq}$, which is a natural consequence of it being generated at a much higher order in the EFT with respect to $c_{qq}$. Therefore, from this type of UV completion presented we expect the scenario in the right panel of Fig.\,\ref{fig:Oh2} to be more naturally realized, where $\Psi\bar\Psi\leftrightarrow q\bar q$ dominates the production but with the BNV still being very relevant due to the decay of the spin-3/2.

One could take one step further in the EFT matching and also consider operators generated at one-loop order, or consider more interactions in Eq.\,\eqref{eq:LUV} such as ones involving the new scalars and the Higgs. Since the number of operators is significantly larger, including many operators involving only SM fields, a general analysis becomes more challenging and is therefore beyond the scope of the present study.

To conclude this Section, we comment on the results obtained recently in Refs.\,\cite{Bellazzini:2025shd,Gherghetta:2025tlx}, which derived constraints on the EFT for spin-3/2 particles based on the consistency of scattering amplitudes. These references conclude that for a theory of a spin-3/2 particle that can be defined for energies $m_\Psi\ll E \ll \Lambda$, the only way for the theory to be non-trivial is to identify $\Psi_\mu$ with the gravitino. With the assumption of the spin-3/2 state being composite, we can at least in a first approximation evade these arguments, as such theories are not defined for $E\gg m_\Psi$; once the energy $E$ is much larger than the confinement scale of the dark gauge group, the description of the theory in terms of dark hadrons break down and we go back to having fundamental dark quarks. In connection to the results obtained in Sec.\,\ref{sec:DM}, we can consistently perform the evolution of DM in the early universe for reheating temperature below its mass $m_\Psi\gtrsim T_\text{RH}$. This region is delimited by the red line in Fig.\,\ref{fig:Oh2}. In the left panel of the figure that assumes only the BNV operator, there is no region where DM can be reproduced while simultaneously satisfying experimental constraints (mainly indirect detection) and $m_\Psi>T_\text{RH}$, therefore disfavouring this scenario completely. For the situation in the right panel, the region of Boltzmann suppressed freeze-in can still be able to reproduce the DM depending on the value of the BNV coupling $\cB$.

%%%%%%%%%%%%%%%%%%%%%%%%%%%%
%%%%% CONCLUSIONS %%%%%%%%%%
%%%%%%%%%%%%%%%%%%%%%%%%%%%%

\section{Conclusions}\label{sec:conclusions}

In this work we have explored a largely uncharted corner of dark matter (DM) phenomenology: a non-supersymmetric spin-$3/2$ particle with baryon number violating (BNV) interactions. Our analysis shows that, despite well-known theoretical challenges associated with elementary spin-$3/2$ fields, a consistent and phenomenologically viable framework can be constructed when these states are treated within an effective field theory (EFT) and, in particular, when they arise as composite states of a confining dark sector.

A central result of the paper is the identification and study of the leading BNV operator at dimension six, which provides a qualitatively new portal between the Standard Model and spin-$3/2$ dark matter. This interaction induces both DM production and decay.  On the cosmological side, we have shown that the observed relic abundance can be achieved through multiple regimes: UV freeze-in, Boltzmann-suppressed freeze-in, and freeze-out. A key qualitative feature is the competition between single-$\Psi$ production, driven by BNV interactions, and pair production mediated by baryon-conserving operators, which scale differently with energy and model parameters. In particular, for generic Wilson coefficients, pair production tends to dominate in the standard UV freeze-in regime.

From the phenomenological perspective, we find a strong complementarity among different experimental probes (see Fig.\,\ref{fig:Oh2}). Indirect detection provides the most stringent constraints in scenarios where the same operator controls both production and decay, with current bounds on the dark matter lifetime $\tau_{\rm DM} \gtrsim 10^{30}~\mathrm{s}$ excluding large portions of parameter space:  $\Lambda/\sqrt{|\cB|}\lesssim 10^{14}$ GeV. Direct detection becomes relevant when baryon-conserving operators govern the scattering with nuclei, probing parts of both the freeze-in and freeze-out regions, more precisely excluding values $\Lambda/\sqrt{|c_{qq}|}\lesssim 10^{5}$ GeV for masses $10~\text{GeV}\lesssim m_\Psi\lesssim 10^4$ GeV. Collider searches, in particular monojet signatures, constrain low-scale interactions and offer sensitivity complementary to direct detection, especially at low dark matter masses $m_\Psi\lesssim 10^3$ (10) GeV for the baryon number conserving (BNV) interactions. The complementarity among theoretical and experimental bounds completely excludes the freeze-out regime for both operators across the viable parameter space.

On the theoretical side, we have demonstrated that a simple dark QCD-like UV completion can naturally realise the scenario. In this construction, the spin-$3/2$ particle emerges as the lightest baryon of a confining $SU(3)_\text{dc}$ dark sector. This setup avoids the main theoretical obstacles typically associated with elementary spin-$3/2$ states, such as issues with superluminal propagation and the consistency of effective interactions, while also preventing catastrophic gravitational production in the early universe. At the same time, it provides a concrete origin for the effective operators considered in our analysis.

Our results open several directions for future work. On the theoretical side, it would be important to further investigate the consistency of interacting spin-$3/2$ EFTs and their embedding in ultraviolet-complete theories. On the phenomenological side, a more refined treatment of hadronic decays at low masses, improved indirect detection analyses, and dedicated collider studies would sharpen the constraints. Finally, exploring the possible connection between BNV spin-$3/2$ dynamics and the origin of the baryon asymmetry remains an especially promising avenue.

In summary, spin-$3/2$ DM with baryonic interactions provides a well-motivated and testable framework that connects strongly coupled dynamics, higher-spin field theory, and experimental searches in a novel way.

%%%%%%%%%%%%%%%%%%%%%%%%%%%%
%%%%% ACKNOWLEDGMENTS %%%%%%
%%%%%%%%%%%%%%%%%%%%%%%%%%%%

\subsection*{Acknowledgments}

We thank Matheus Martines, Luighi P. S. Leal, Gabriele Levati, Fabian Esser, and Pedro Bittar for useful discussions, and especially Bruno Siqueira Eduardo not only for many discussions but also for cross-checking some results with Hilbert Series techniques. We are also deeply grateful to Laura Covi and Enrico Bertuzzo for discussions and thoughtful comments on the manuscript.\\

FC acknowledges partial support from the FORTE project CZ.02.01.01/00/22 008/0004632 co-funded by the EU and the Ministry of Education, Youth and Sports of the Czech Republic. FC acknowledges partial financial support from the Science Foundation Ireland Grant 21/PATHS/9475 (MOREHIGGS) under the SFI-IRC Pathway Programme. GMS acknowledges financial support from the Deutsche Forschungsgemeinschaft under Germany’s Excellence Strategy – EXC 2121
“Quantum Universe”, as well as from the grant 491245950. This project received funding from the European Union’s Horizon Europe research and innovation
program under the Marie Sk\l odowska-Curie Staff Exchange grant agreement No 101086085
– ASYMMETRY.\\

\appendix

\section{Spin 3/2 basics}
\label{app:spin32}

Let us review the QFT framework describing spin-3/2 particles, namely the Rarita--Schwinger Lagrangian\,\cite{Rarita:1941mf,Pascalutsa:2000kd,Pilling:2004cu}. It is based on the observation that, in terms of $SU(2)\times SU(2)$ classification of the Lorentz group, we can obtain a spin-3/2 representation by composing vector and spinor representations
\al{\label{eq:RS_decomposition}
&\Psi_\mu\sim \left(\frac{1}{2},\frac{1}{2}\right)\otimes\left[\left(0,\frac{1}{2}\right)\oplus\left(\frac{1}{2},0\right)\right]  \\ 
&\hspace{1cm}= \left(1,\frac{1}{2}\right)\oplus\left(\frac{1}{2},1\right)\oplus\left(0,\frac{1}{2}\right)\oplus\left(\frac{1}{2},0\right),
}
where we denote the Rarita--Schwinger field as $\Psi_\mu$. In the right-hand side of the formula above, the first half contains the helicity modes $\pm 3/2$ and $\pm1/2$, whereas the second half is a pure spin-$1/2$ piece that needs to be factorized out. By imposing the conditions
\be\label{eq:RS_constraints}
\gamma^\mu\Psi_\mu=0,\quad \del^\mu\Psi_\mu=0,
\ee
one can remove the spurious degrees of freedom and is left with only the physical degrees of freedom for a massive spin-3/2 particle, as we will see in the following.

The spurious spin-1/2 component in the decomposition of Eq.\,\eqref{eq:RS_decomposition} can be removed by requiring the theory to be invariant under the following transformation
\be\label{eq:Psi_SUSY_transformation}
\Psi_\mu \to \Psi_\mu +\del_\mu \epsilon,
\ee
where $\epsilon$ is a local spinor. From the transformation in Eq.\,\eqref{eq:Psi_SUSY_transformation} we can determine the correct kinetic Lagrangian. Its most general form is
\be
\La_\text{RS} \supset \bar \Psi_\mu \Gamma^{\mu\nu\alpha}\del_\alpha \Psi_\nu .
\ee
The quantity $\Gamma^{\mu\nu\alpha}$ is in general a combination of $\gamma$-matrices and, because the field contains spin-1/2 components and must thus have the same dimension as a Dirac spinor, we factored a derivative out of it. Once we apply Eq.\,\eqref{eq:Psi_SUSY_transformation} and imposing $\delta \La_\text{RS}=0$ for any $\epsilon$, we conclude that $\Gamma^{\mu\nu\alpha}$ is fully anti-symmetric. In particular, it means that it is proportional to
\be\label{eq:Gamma_munualpha}
\Gamma^{\mu\nu\alpha}\propto \gamma^{\mu\nu\alpha} \equiv i\epsilon^{\mu\nu\alpha\beta}\gamma^5\gamma_\beta.
\ee
The proportionality constants are determined by canonically normalizing the Hamiltonian, leading to
\be\label{eq:RaritaSchwinger_massless}
\La_\text{RS}\supset \epsilon^{\mu\nu\alpha\beta} \bar\Psi_\mu \gamma^5 \gamma_\alpha \del_\beta \Psi_\nu.
\ee
In the massless case the second spin-1/2 component, which correspond to the longitudinal modes in the massive case, are also unphysical. These can be removed by fixing that $\gamma^\mu\Psi_\mu=0$. Then, computing the equations of motion from Eq.\,\eqref{eq:RaritaSchwinger_massless} and imposing this condition we find $\Psi_\mu$ needs to satisy in addition: $\del^\mu\Psi_\mu =0$. We therefore find that the constraints in Eq.\,\eqref{eq:RS_constraints} plus the transformation in Eq.\,\eqref{eq:Psi_SUSY_transformation} eliminate all unphysical degrees of freedom in the massless scenario.

If we now give a mass to the Rarita--Schwinger field, the mass term is in general given by
\be
\bar \Psi_\mu M^{\mu\nu} \Psi_\nu,
\ee
which can be expressed as $M^{\mu\nu} = a\,\eta^{\mu\nu}+ib \,\sigma^{\mu\nu}$, $a,b\in \mathbb{C}$ and $\sigma^{\mu\nu}=i[\gamma^\mu,\gamma^\nu]/4$. To determine this term, we can extract the spurious spin-1/2 component using Eq.\,\eqref{eq:Psi_SUSY_transformation} and find the necessary constraints on $\Psi_\mu$ such to have the variation of the Lagrangian to be zero. Doing so implies that $a=0$ and the conditions in Eq.\,\eqref{eq:RS_constraints} need to be satisfied. The massive free Rarita--Schwinger Lagrangian is thus
\be\label{eq:RaritaSchwinger_massive}
\La_\text{RS} = \epsilon^{\mu\nu\alpha\beta} \bar\Psi_\mu \gamma^5 \gamma_\alpha \del_\beta \Psi_\nu - 2im_\Psi\bar \Psi_\mu \sigma^{\mu\nu}\Psi_\nu,
\ee
and the field satisfies
\be
\gamma^\mu\Psi_\mu = 0,\quad \del^\mu \Psi_\mu =0.
\ee
Using the equation above, the equations of motion become
\be\label{eq:RS_EoM}
(i\slashed{\del}-m_\Psi)\Psi_\mu = 0,
\ee
which is nothing but a Dirac equation for each of the vector components of the field. Finally, the propagator of the massive Rarita--Schwinger field is
\begin{align}\label{eq:RS_propagator}
    S^{\mu\nu}(p) =& \frac{\slashed{p}+m_\Psi}{p^2-m^2_\Psi}\left[-\eta^{\mu\nu} + \frac{1}{3}\gamma^\mu\gamma^\nu \right.\\  +&\frac{1}{3m_\Psi}\left(\gamma^\mu p^\nu - \gamma^\nu p^\mu\right)+\frac{2}{3m_\Psi^2}p^\mu  p^\nu  \biggl], \nonumber
\end{align}
from which we automatically get the spin sum for the wave-functions $u_\mu(p)$. Another possible representation of the spin-3/2 polarization sum is given by the William's representation:
\begin{align}
P_{\mu\nu}(p) &= -(\slashed{p}-m_\Psi)\left[\Pi_{\mu\nu}(p)-\frac{1}{3}\Pi_{\mu\alpha}(p)\Pi_{\nu\beta}(p)\gamma^\alpha\gamma^\beta\right], \nonumber\\ &\Pi_{\mu\nu}(p)=\eta_{\mu\nu}-\frac{p_\mu p_\nu}{m_\Psi^2},
\end{align}
that coincides with Eq.\,\eqref{eq:RS_propagator} when the spin-3/2 is on-shell.

\section{Baryon number violating operators}
\label{app:misc}

In the main text we have argued that the dimension six operator in Eq.\,\eqref{eq:La_dim6_BNV} is the sole operator at this mass dimension that violates baryon number. Nevertheless, we could have in principle another operator, namely~\footnote{The BNV operator in Eq.\,\eqref{eq:La_dim6_BNV} can be brought to similar form using Fierz identities\,\cite{Liao:2012uj}.}
\be\label{eq:udd}
\La_\text{BNV} = \frac{c_{udd}^{ijk}}{\Lambda^2}(\bar \Psi_\mu \gamma^\nu u_{Ri})(\bar d_{Rj}^c\sigma^{\mu\nu} d_{Rk})+h.c.,
\ee
where the flavor structure of the $d_R$ pair must be symmetric, $c_{udd}^{ijk}=c_{udd}^{jik}$. To understand why this operator does not contribute, we can use on-shell methods (see for instance Refs.\,\cite{Dixon:2013uaa,Elvang:2013cua,Arkani-Hamed:2017jhn,Travaglini:2022uwo, Henn:2014yza}). Using the notation of Refs.\,\cite{Gherghetta:2024tob,Gherghetta:2025tlx} for the massive spin-3/2 spinor variables, the amplitude for this vertex is written as
\begin{align}
\Amp[\Psi_1 u_{R2}d_{R3}d_{R4}] 
& \propto \frac{1}{\Lambda^2} \asbraket{\bb{1}}{\gamma^\nu}{2} \frac{\asbraket{\bb{1}}{\gamma^\mu}{\bb{1}}}{\sqrt{2}m_\Psi} \sbraket{3|\sigma^{\mu\nu} |4} \nonumber \\
& \propto \frac{1}{\Lambda^2} \epsilon^{IJ} \sbraket{1^K2}\sbraket{34}+(\text{symm.}~IJK)\nonumber \\
& = 0 ,
\end{align}
where the numbers are the labels for momenta, we have used Shouten's identity and the fact that the amplitude is symmetrised in the little-group indices $IJK$ of each particle. The resulting amplitude is proportional to $\epsilon^{IJ}$, meaning that it transforms trivially under these little-group indices and consequentely vanishing given the symmetrisation over the litte-group indices. This means that the operator in Eq.\,\eqref{eq:udd} will not contribute to any of the scattering processes considered in this work.

Contrary to Eq.\,\eqref{eq:udd}, the operator in Eq.\,\eqref{eq:La_dim6_BNV} does not behave so. The amplitude is
\be
\Amp[\Psi_1 Q_2 Q_3 d_{R4}] \propto \frac{1}{m_\Psi\Lambda^2} \braket{\bb{1}2}\braket{\bb{1}{3}} \sbraket{\bb{1}4},
\ee
which $(i)$ is what one would expect from EFT arguments and $(ii)$ is explicilty dependent on the spin-1/2 helicities of the spin-3/2 particle, thus having a problematic behaviour with $m_\Psi\to 0$.

\section{Boltzmann Equation}
\label{app:boltzmannEq}

Let us introduce the Boltzmann equation for $1+2 \rightarrow a+b$ and $1\rightarrow a+b+ \dots  n$ processes to study the evolution of the DM number density. It reads
\al{
    H S x_1 \frac{\mathrm{d} Y_1}{\mathrm{d}x_1} &= \dot n_1 + 3Hn_1 \\=& \gamma_{12\leftrightarrow ab} \left( \frac{Y_a Y_b}{Y^{\rm eq}_a Y^{\rm eq}_b} - \frac{Y_1 Y_2}{Y^{\rm eq}_1 Y^{\rm eq}_2} \right) 
    \\ +& \gamma_{1 \leftrightarrow ab \dots n} \left( \frac{Y_a Y_b \dots Y_n}{Y^{\rm eq}_a Y^{\rm eq}_b \dots Y^{\rm eq}_n} - \frac{Y_1}{Y^{\rm eq}_1} \right) \, ,
}
where $x=m/T$, $H$ the Hubble rate, $S$ is the entropy density and $Y_1=n_1/S$ is the yield.
To obtain the above equation, we used the Maxwell-Boltzmann approximation for the equilibrium distribution $f^{\rm eq} \sim e^{-E/T}$, imposed $CP$ invariance and neglected the Bose ehancement and Pauli blocking terms\,\cite{Kolb:1990vq}. For the reaction rate of a decay process we have
\begin{equation}
\gamma_{1 \leftrightarrow ab \dots n}
= n_1^{\mathrm{eq}} \langle \Gamma_{1 \to ab \dots n} \rangle
= n_1^{\mathrm{eq}} \frac{K_1(x)}{K_2(x)} \, \Gamma_{1 \to ab \dots n} \,,
\end{equation}
where $\Gamma_{1 \to ab \dots n}$ is the standard QFT decay width, and $K_j$ is the modified Bessel function of the second kind of order $j$. The reaction rate of the process $\gamma_{12\leftrightarrow ab} $ is defined as
\begin{align}
\gamma_{1+2 \leftrightarrow a+b} 
&= n_1^{\rm eq} n_2^{\rm eq} \langle \sigma_{1+2 \leftrightarrow a+b} v \rangle \\
&= \frac{T}{64\pi^4} \int_{s_{\min}}^{\infty} ds \, \sqrt{s}\, K_1\!\left(\frac{\sqrt{s}}{T}\right) \, \hat{\sigma}_{1+2 \leftrightarrow a+b}(s) \nonumber,
\end{align}
where $s_{\min} = \max\!\left[(m_1 + m_2)^2,\,(m_a + m_b)^2\right]$ and $\langle \sigma_{1+2 \leftrightarrow a+b} v \rangle$ is the thermally averaged cross section. The modified cross-section is 
\begin{align}
\hat{\sigma}_{1+2 \leftrightarrow a+b}(s) 
= \frac{g_1 g_2}{c_{12}} \frac{2\,\lambda(s, m_1^2, m_2^2)}{s} \, \sigma_{1+2 \leftrightarrow a+b}(s) .
\end{align}
where the K\"allen function is $\lambda(x,y,z) = (x - y - z)^2 - 4yz$ and $\sigma_{1+2 \leftrightarrow a+b}(s)$ is the standard cross-section, averaged over the initial state and summed over the final ones. The factor $c_{12}$ is equal to 2 for identical initial state particles and to 1 otherwise. Let us now discuss the approximations used respectively in the freeze-in and freeze-out scenarios.

\subsection{Freeze-in approximations}

Since the non-renormalisable freeze-in is a UV dominated process\,\cite{Elahi:2014fsa}, we can neglect the spin-$3/2$ mass appearing in the lower limit of the $s$ integral and in the K\"allen functions, where we have terms like $\sqrt{s- m_\Psi^2} \simeq s$. Moreover, taking the $s\gg m_\Psi^2$ limit, the cross sections are dominated by term with the highest power of $s$ and their expression are (see Eq.\,\eqref{eq:sigma_FI})
\al{
\sigma_{\bar q\Psi\to qq}(s) &\simeq\frac{|\cB|^2 N_f(s)}{72\pi m_\Psi^2 \Lambda^4} s^2,\\
\sigma_{ \Psi \bar\Psi\to  q\bar q}(s)&\simeq\frac{|c_{qq}|^2N_f'(s)}{6912 \pi m_\Psi^4\Lambda^4}s^3.
}
Therefore the reaction rate can be analytically computed and the generic solution for a $s^n$ cross-section scaling, is given by
\begin{align}
\label{eq:approx_reaction_rates}
    \gamma\sim T\int_0^\infty\dd s\ s^n \sqrt{s} s& K_1(\sqrt{s}/T)  \\
    &= 4^{2+n}\Gamma(2+n)\Gamma(3+n) T^{6+2n},\nonumber
\end{align}
which for the two operators we considered is
\al{
\gamma_{\bar q\Psi\leftrightarrow qq} &\simeq\frac{384 |\cB|^2 N_f}{\pi^5 m_\Psi^2 \Lambda^4} T^{10},\\
\gamma_{ \Psi \bar\Psi\leftrightarrow  q\bar q}&\simeq\frac{640 |c_{qq}|^2 N_f'}{3 \pi^5 m_\Psi^4\Lambda^4}T^{12}.
}
Notice that extracting the $N_f$ ($N_f'$) from the integral is an approximation. It is justified since they are approximately piecewise functions of $s$. With the approximation of the reaction rate we can then compute the yield integrating the Boltzmann equation in the temperature between $T \rightarrow 0$ and $T \rightarrow \infty$. 
\be\label{eq:Boltzmann_FI}
Y_\Psi \simeq  \int^{\infty}_{0} \dd x \; \frac{\gamma_{\bar q\Psi \leftrightarrow qq} + \gamma_{\Psi\bar\Psi \leftrightarrow q\bar q}}{H S x},
\ee
The final results are presented in Sec.\,\ref{sec:DM}.

\subsection{Freeze-out approximations}

Let us discuss now the approximation needed in the freeze-out case.
Considering the DM freezing out when non-relativistic ($x \gg 1$), we have two contributions to the thermally averaged cross-section integrand that change rapidly with $s$, namely the Bessel function $K_1(\sqrt{s}/T)$ and the K\"allen functions. The Bessel function is exponentially suppressed for $\sqrt{s} < T$, while the K\"allen functions
\al{\label{eq:Kallen}
    \lambda(s,m_\Psi^2,0) &= (s-m_\Psi^2)^2 ,\qquad \mathrm{for } \;\; \bar q \Psi\leftrightarrow qq,\\
    \lambda(s,m_\Psi^2,m_\Psi^2) &= (s-4m_\Psi^2)^2,  \qquad \mathrm{for } \;\; \Psi\bar  \Psi\leftrightarrow q \bar q ,
}
go to zero at $s^* = m_\Psi^2$ ($4m_\Psi^2$). Therefore, we expect the cross section to have a maximum close to $s \sim s^*$. We then approximate the terms slowly varying in $s$ as constant and we evaluate them at $s^*$. Doing so we obtain the maximum of the $\langle \sigma v\rangle$ integrands to be 
\begin{align}\label{eq:smax}
    &\frac{s_\text{max}}{T^2} \simeq 8 + x^2 + 4\sqrt{4+x^2}, \quad \mathrm{for } \;\; \bar q \Psi\leftrightarrow qq,\\
   &\frac{\sqrt{s_\text{max}}}{T}\simeq \frac{17}{16}+\frac{1}{16}\sqrt{463 + 1024 x^2},  \quad \mathrm{for } \;\; \Psi\bar  \Psi\leftrightarrow q \bar q .  \nonumber
\end{align}
We can now find an analytical approximation of the interaction rate for the two operators. For example, the terms coming from the Källen functions in Eq.\,\eqref{eq:Kallen} can be approximated using Eq.\,\eqref{eq:smax} to leading order as
\al{
s_\text{max}-m_\Psi^2 &\simeq 4m_\Psi T , \qquad \mathrm{for }  \;\; \bar q \Psi\leftrightarrow qq,\\
s_\text{max}-4 m_\Psi^2&\simeq 5m_\Psi T,  \qquad \mathrm{for }  \;\; \Psi\bar  \Psi\leftrightarrow q \bar q .
}
The interaction rates, defined in Eq.~\eqref{eq:FO_rates_approx}, in the freeze-out approximation, are thus given by
\begin{align}
\Gamma_{\Psi q qq} 
& = \frac{3}{4\pi^4}\frac{T}{n_\Psi^\text{eq}}\int_{m_\Psi^2}^\infty \dd s~\sigma_{\bar q\Psi\to qq}(s)\frac{(s-m_\Psi^2)^2}{\sqrt{s}}K_1(\sqrt{s}/T) \nonumber \\
& \simeq \frac{8N_f |c_{\slashed{B}}|^2m_\Psi^5}{e^2\pi^3\Lambda^4}\frac{1}{x^3}\,,
\end{align}
for $\bar{q}\Psi \leftrightarrow qq$ and 
\be
\Gamma_{\Psi\Psi qq} \simeq \frac{85 \sqrt{17}e^{-17/16}}{73728\sqrt{2}\pi^3}\frac{|c_{qq}|^2N_f'(m_\Psi)^2m_\Psi^5e^{-x}}{x^{3/2}\Lambda^4}\,,
\ee
for $\Psi \bar \Psi \leftrightarrow q \bar q$. Since the integrand is strongly peaked at $s_{\rm max}$, we approximated the integral as twice the rectangle in the interval $(s_\text{max}-m_\Psi^2)$ or $(s_\text{max}-4m_\Psi^2)$ accordingly, with the integrand being evaluated at $s_\text{max}$. Notice that in the Bessel function approximation we include a subleading term because it leads to a relevant numerical factor in order to match the numerical results. From the interaction rate one can compute the decoupling and freeze-out temperature, as shown in Sec.\,\ref{sec:DM}. The yield for the $\Psi \bar \Psi \leftrightarrow q \bar q$ is then obtained from the standard freeze-out derivation as\,\cite{Kolb:1990vq}
\al{\label{eq:Yqq_FO_apprx}
\frac{\dd Y_\Psi}{Y_\Psi^2} \simeq &- \frac{\braket{\sigma_{\Psi\bar\Psi\rightarrow q\bar q} v} S}{x H}\Bigg|_{x=1} \frac{\dd x}{x^2} \\  \Rightarrow & Y_\Psi(\infty) \simeq x_\text{FO}\frac{x H}{\braket{\sigma_{\Psi\bar\Psi\rightarrow q\bar q} v} S}\Bigg|_{x=1},
}
with $x_\text{FO}$ being computed in Eq.\,\eqref{eq:x_FO_estimate}, defined as the moment of (very fast) freeze-out according to Eq.\,\eqref{eq:xFO_def}. While, for $\bar{q}\Psi \leftrightarrow qq$ we have a slower freeze-out since the interaction rate is not exponentially suppressed and we can calculate the yield as 
\al{\label{eq:Yqqq_FO_approx}
xH \frac{\dd Y_\Psi}{\dd x} \simeq & - \Gamma_{\Psi qqq}(x) Y_\Psi \\ &\Rightarrow \int_{Y_\Psi( x_{\rm dec})}^{Y_\Psi(\infty)}\frac{\dd Y_\Psi}{Y_\Psi} \simeq - x_\text{\rm dec}^2\int_{\bar x}^\infty \frac{\dd x}{x^2}   \\ & \Rightarrow Y_\Psi(\infty) \simeq Y_\Psi^{\rm eq}( x_{\rm dec})e^{-x_\text{dec}},
}
where again $x_\text{dec}$ is given in Eq.\,\eqref{eq:x_FO_estimate} and previously defined in Eq.\,\eqref{eq:xdec_def} for a scenario of slow freeze-out.

% \newpage
\pagestyle{plain}
\bibliographystyle{JHEP2}
\small
\bibliography{draft.bib}
\end{document}